\documentclass[lettersize,journal]{IEEEtran}

\usepackage{amsmath}
\usepackage{amsfonts}
\usepackage{amssymb}
\usepackage{mathtools}
\usepackage{bm}

\usepackage{algorithm}
\usepackage{algorithmic}

\usepackage{array}
\usepackage{booktabs}
\usepackage{multirow}
\usepackage{makecell}
\usepackage{tabularx}
\usepackage{threeparttable}

\usepackage{graphicx}
\usepackage{caption}
\usepackage{subcaption}
\usepackage{dblfloatfix}
\usepackage{float}

\usepackage{amsthm}

\newtheorem{theorem}{Theorem}
\newtheorem{lemma}{Lemma}

\theoremstyle{plain}
\newtheorem{proposition}{Proposition}
\usepackage{textcomp}
\usepackage{url}
\usepackage{verbatim}
\usepackage{cite}
\usepackage{rotating}

\usepackage{xcolor}
\usepackage{siunitx}
\usepackage{microtype}
\usepackage{amsmath}

\usepackage{booktabs}    
\usepackage{siunitx}     
\usepackage{colortbl}   
\usepackage{xcolor}     
\usepackage{array}
\newcolumntype{P}[1]{>{\raggedright\arraybackslash}p{#1}}
\usepackage{pifont}
\newcommand{\xmark}{\ding{55}} 
\usepackage{pifont}
\newcommand{\cmark}{\ding{51}} 

\usepackage[hidelinks]{hyperref}

\begin{document}

\title{EdgeDetect: Importance-Aware Gradient Compression with Homomorphic Aggregation for Federated Intrusion Detection}

\author{Noor Islam S. Mohammad%
\thanks{Department of Computer Science, Istanbul Technical University, Maslak, TR (Corresponding author: islam23@itu.edu.tr).}%
\thanks{This research received no external funding.}
}

\markboth{EdgeDetect: Secure Gradient Compression for Federated Intrusion Detection}%
{Mohammad: Binarized Gradient Aggregation for Edge Intrusion Detection}

\maketitle

\begin{abstract}
Federated learning (FL) enables collaborative intrusion detection without raw data exchange, but conventional FL incurs high communication overhead from full-precision gradient transmission and remains vulnerable to gradient inference attacks. This paper presents \textbf{EdgeDetect}, a communication-efficient and privacy-aware federated IDS for bandwidth-constrained 6G-IoT environments. EdgeDetect introduces \emph{gradient smartification}, a median-based statistical binarization that compresses local updates to $\{+1,-1\}$ representations, reducing uplink payload by $32\times$ while preserving convergence. We further integrate Paillier homomorphic encryption over binarized gradients, protecting against honest-but-curious servers without exposing individual updates. Experiments on CIC-IDS2017 (2.8M flows, 7 attack classes) demonstrate $98.0\%$ multi-class accuracy and $97.9\%$ macro F1-score, matching centralized baselines, while reducing per-round communication from $450$~MB to $14$~MB ($96.9\%$ reduction). Raspberry Pi~4 deployment confirms edge feasibility: $4.2$~MB memory, $0.8$~ms latency, and $12$~mJ per inference with $<0.5\%$ accuracy loss. Under $5\%$ poisoning attacks and severe imbalance, EdgeDetect maintains $87\%$ accuracy and $0.95$ minority class F1 ($p<0.001$), establishing a practical accuracy–communication–privacy tradeoff for next-generation edge intrusion detection.
\end{abstract}


\section{Introduction}

\IEEEPARstart{N}{ext-generation} Wireless technologies 5G, 6G, and IoT enable massive machine-type communications and ultra-reliable low-latency services \cite{mcmahan2017communication, liu2023feddnn} while simultaneously expanding the attack surface for sophisticated cyber threats. As billions of heterogeneous edge devices generate high-volume traffic in smart cities, autonomous vehicles, and Industry 4.0, traditional centralized IDS face fundamental limitations, including scalability bottlenecks, communication latency, single points of failure, and difficulty handling high dimensionality and severe class imbalance in modern network traffic.

Machine learning has become central to automated threat identification \cite{yang2019federated, mothukuri2021federated}. However, centralized deployment of deep learning architectures requires aggregating raw sensor readings, enterprise logs, and user data at cloud servers, exposing systems to potential data breaches and regulatory violations \cite{nguyen2022federated}. Federated Learning (FL) addresses this limitation by enabling collaborative model training while preserving data locality \cite{yin2023comprehensive}. 

Despite its advantages, practical FL implementations face two critical challenges: (1) \emph{communication overhead}, transmitting high-dimensional gradient vectors from thousands of edge clients consumes excessive bandwidth; and (2) \emph{gradient leakage},  shared model updates may be reverse-engineered to reconstruct sensitive training samples \cite{siriwardhana2021federated}.  

To address these challenges, we propose EdgeDetect, a scalable and privacy-aware federated IDS tailored for resource-constrained 6G-IoT environments \cite{zhou2019edge}. EdgeDetect introduces a novel \emph{gradient smartification} mechanism that transforms continuous gradient updates into lightweight binarized representations ($\{+1,-1\}$) using \emph{median-based statistical thresholding}. This adaptive, distribution-aware compression reduces uplink payload size by up to $32\times$ while preserving empirical convergence behavior \cite{haddadpour2021federated, lim2020federated}. 

Unlike fixed-threshold methods (e.g., signSGD), our approach suppresses low-magnitude gradient components below the per-client median, reducing stochastic noise and improving stability under heterogeneous data distributions. We further integrate Paillier homomorphic encryption over the binarized gradients, ensuring that only aggregated model updates are visible to the central server, providing strong cryptographic protection against gradient inversion and honest-but-curious adversaries \cite{lim2020federated}. The joint optimization of compression and privacy enables EdgeDetect to achieve both communication efficiency and end-to-end confidentiality without compromising detection accuracy.

\subsection{Contribution}
This work makes the following key contributions:

\begin{itemize}
\item Alignment-Aware Federated IDS Architecture: We present EdgeDetect, a privacy-preserving federated intrusion detection framework designed for 6G-IoT environments~\cite{chen2024privacy}. The architecture integrates PCA-based dimensionality reduction, imbalance-aware sampling, and secure aggregation within a unified decentralized pipeline, enabling collaborative learning without sharing raw network traffic while maintaining scalability and robustness.

\item Adaptive Median-Based Gradient Smartification with Encrypted Aggregation: We introduce a statistically adaptive median-threshold binarization strategy that compresses gradients into $\{+1,-1\}$ while preserving directional alignment under heterogeneous and heavy-tailed client distributions. In contrast to fixed zero-threshold signSGD~\cite{bernstein2018signsgd}, the proposed per-client adaptive rule improves convergence stability. Combined with Paillier homomorphic encryption applied directly to binarized gradients, the method achieves up to $32\times$ communication reduction while mitigating gradient inversion risks~\cite{chen2020convergence,aouedi2022federated}.

\item Quantified Privacy--Utility--Efficiency Trade-off: Extensive ablation and adversarial analyses demonstrate $98.0\%$ multi-class accuracy with $96.9\%$ communication reduction on CIC-IDS2017 (2.8M flows), achieving performance comparable to centralized baselines while providing cryptographic privacy guarantees. The framework maintains $>85\%$ accuracy with 20\% malicious clients and reduces inversion PSNR from 31.7~dB to 15.1~dB~\cite{mcmahan2017communication}.

\item Edge-Validated Deployment: Real-world deployment on Raspberry Pi~4 devices confirms practical feasibility, requiring only 4.2~MB memory, 0.8~ms latency, and 12~mJ per inference, with less than $0.5\%$ accuracy degradation. These results validate suitability for resource-constrained 6G-IoT edge environments.
\end{itemize}

\section{Related Work}

The evolution of IDS from signature-based systems to ML and deep learning paradigms has significantly advanced network security \cite{wei2020federated, abadi2016deep}. This section reviews anomaly detection in wireless networks, FL for decentralized security, and privacy–communication efficiency challenges.

\subsection{Deep Learning-Based Anomaly Detection}

Deep learning has become the standard for detecting complex attack patterns in high-dimensional network traffic. Classical algorithms such as SVMs and random forests remain competitive for structured features \cite{ma2022privacy, truex2019hybrid}, while CNN–RNN and LSTM architectures capture temporal dependencies for DDoS and zero-day detection \cite{chen2024privacy}. Image-based encodings of time-series traffic further enhance spatial feature extraction \cite{zhang2024secure}. However, these centralized approaches require large-scale data aggregation, introducing privacy risks and system-level vulnerabilities.

\subsection{Federated Learning in IoT Networks}

FL enables decentralized training without sharing raw data \cite{bonawitz2017practical}. Applications include IoT security, industrial sensor networks, and cross-domain intrusion detection \cite{rahman2020internet}. Edge–cloud collaborative architectures reduce response latency while preserving data locality \cite{bell2020secure, kadhe2020fastsecagg}. However, standard FL algorithms such as FedAvg rely on full-precision gradient exchange, creating communication bottlenecks in bandwidth-limited 6G IoT systems \cite{zhang2020batchcrypt, alistarh2017qsgd}.

\subsection{Privacy Preservation and Gradient Compression}

While FL mitigates raw data exposure, it remains vulnerable to gradient inference attacks. Differential Privacy (DP) and Homomorphic Encryption (HE) improve confidentiality but may introduce accuracy or computational overhead \cite{fang2022lightsecagg, so2021turbo}. Communication-efficient methods such as signSGD and gradient sparsification reduce bandwidth requirements \cite{bernstein2018signsgd, lin2018deep}. 

However, few approaches jointly optimize gradient compression and encrypted aggregation in resource-constrained intrusion detection settings. Our PoL-based gradient smartification mechanism integrates statistical binarization with encrypted aggregation to address both communication efficiency and privacy preservation \cite{reisizadeh2020fedpaq, sattler2019robust, ma2024efficient}.

\subsection{Distinction from signSGD and Quantized FL}
Unlike fixed-threshold quantizers such as QSGD~\cite{alistarh2017qsgd} or TernGrad~\cite{wen2017terngrad}, which apply uniform quantization levels, our median-threshold binarization adapts to the per-client gradient distribution. This property is especially valuable for IDS data, where gradients exhibit heavy tails due to rare attack events.

\begin{equation}
    \tau_t = \text{median}(g_t)
\end{equation}

Thus, smartification preserves relative ordering information within each gradient vector and adapts to heavy-tailed feature distributions typical in IDS models. Distribution-adaptive quantization with provable descent guarantees and entropy-aware privacy strengthening.

\section{System Architecture (IDS)}

We propose \textit{EdgeDetect}, a privacy-preserving federated learning architecture for 6G-enabled IoT environments \cite{ma2024efficient, preuveneers2018distributed, zhao2020intelligent}. The system comprises $K$ resource-constrained edge clients and a central aggregation server collaboratively training a global anomaly detection model $M_{\text{global}}$ without exposing private local datasets $\mathcal{D}_i$.  

\textit{Each communication round consists of four phases:}
\subsubsection{Phase 1: Client-Side Local Training}

Let's $\mathcal{S} = \{1, 2, ..., K\}$ denote the participating clients. At round $r$, the server broadcasts global parameters $W^{(r)}$ \cite{bonawitz2019towards, li2020federated}. Each client performs $E$ local epochs minimizing $\mathcal{L}(W_i, \mathcal{D}_i)$:

\begin{equation}
W_i^{(r+1)} = W_i^{(r)} - \eta \nabla \mathcal{L}(W_i^{(r)}, \mathcal{D}_i)
\end{equation}

The model update is

\begin{equation}
\Delta_i^{(r)} = W_i^{(r+1)} - W^{(r)}.
\end{equation}

\textbf{Phase 2: Gradient Smartification.} 
To reduce uplink communication cost, we apply a statistical binarization operator \(\Phi(\cdot)\):
\[
\Delta_{i,j}^{\mathrm{bin}} = 
\begin{cases} 
+1, & \text{if } \Delta_{i,j}^{(r)} \geq \theta_i \\
-1, & \text{otherwise}
\end{cases}
\]
where \(\theta_i = \mathrm{median}(|\Delta_i^{(r)}|)\) is the median of the absolute values of the local gradient vector. The resulting vector \(\Delta_i^{\mathrm{bin}} \in \{+1,-1\}^d\) compresses the representation by \(32\times\) while preserving directional information.

\subsubsection{Phase 3: Privacy-Preserving Encryption}

Each client encrypts $\Delta_i^{bin}$ using a homomorphic encryption scheme $\mathcal{E}(\cdot)$ \cite{cheng2021secureboost, fereidooni2021safelearn}:

\begin{equation}
C_i^{(r)} = \mathcal{E}(\Delta_i^{bin})
\end{equation}

ensuring that individual updates remain confidential during transmission.

\subsubsection{Phase 4: Secure Aggregation and Global Update}

Upon receiving ciphertexts from active clients $S_r \subseteq \mathcal{S}$, the server performs encrypted aggregation \cite{tang2019doublesqueeze, haddadpour2021federated}:

\begin{equation}
\Delta_{agg}^{bin} = \frac{1}{|S_r|} \sum_{i \in S_r} \mathcal{D}(C_i^{(r)})
\end{equation}

The global model is updated as

\begin{equation}
W^{(r+1)} = W^{(r)} + \alpha \cdot \Delta_{agg}^{bin}.
\end{equation}

\section{Methodology}

\subsection{Data Exploration and Preprocessing}

The CIC-IDS2017 dataset contains 2,830,743 records with 79 features. Exploratory data analysis revealed:  (i) 308,381 duplicate rows, removed to mitigate potential overfitting bias;  (ii) missing and infinite values in \texttt{Flow Bytes/s} and \texttt{Flow Packets/s} (0.06\%), imputed using median statistics to preserve distributional robustness;  (iii) high memory consumption ($\approx 1.5$ GB), mitigated via numerical downcasting (float64 to $\rightarrow$ float32, int64 to $\rightarrow$ int32), achieving 47.5\% memory reduction; and  (iv) severe class imbalance with benign traffic dominating attack categories. 

To ensure computational feasibility, a 20\% stratified sample was extracted. Statistical validation confirmed representativeness, with feature mean deviations below 5\% relative to the full dataset.

\subsection{Feature Engineering and Selection}

Temporal features (e.g., flow inter-arrival statistics) capture the bursty nature of volumetric attacks, while entropy-based features quantify the randomness in packet sizes, which often deviates during scanning or exfiltration attempts~\cite{rothchild2020fetchsgd, lu2020adaptive}.

\textbf{Temporal Features:} Flow inter-arrival time statistics were computed as
\begin{equation}
\begin{aligned}
\Delta t_{\text{mean}} &= \frac{1}{n-1} \sum_{i=2}^{n} (t_i - t_{i-1}),\\
\Delta t_{\text{std}}  &= \sqrt{\frac{1}{n-1} \sum_{i=2}^{n} (\Delta t_i - \Delta t_{\text{mean}})^2 }.
\end{aligned}
\end{equation}

\textbf{Entropy-Based Features:} Packet size entropy captures distributional randomness \cite{lu2020adaptive, wen2017terngrad}:
\begin{equation}
H(S) = - \sum_{s \in \mathcal{S}} p(s) \log_2 p(s),
\end{equation}
where $\mathcal{S}$ denotes unique packet sizes and $p(s)$ their empirical probabilities.

\textbf{Feature Selection:} Recursive Feature Elimination (RFE) was applied using Random Forest permutation importance \cite{basu2020qsparse, mishchenko2019distributed}:
\begin{equation}
I_j = \frac{1}{T} \sum_{t=1}^{T} \mathbb{I}\!\left( f_t(D) \neq f_t^{(-j)}(D) \right),
\end{equation}
where $f_t^{(-j)}$ denotes a tree $t$ with a feature $j$ permuted. Features were ranked according to $I_j$ and selected prior to dimensionality reduction.

\subsection{Dimensionality Reduction via Incremental PCA}

To mitigate multicollinearity (23\% of feature pairs with $|\rho|>0.8$) and reduce computational overhead, incremental PCA was applied to the standardized feature matrix $Z \in \mathbb{R}^{n \times d}$ ($d=78$) \cite{koloskova2019decentralized, horvath2022natural}:
\begin{equation}
\mathrm{Cov}(Z) = \frac{1}{n-1} Z^\top Z = V \Lambda V^\top.
\end{equation}

The reduced representation was obtained as
\begin{equation}
Z_{\text{PCA}} = Z V_k,
\end{equation}
retaining $k=35$ principal components satisfying
\begin{equation}
\frac{\sum_{i=1}^{k} \lambda_i}{\sum_{i=1}^{d} \lambda_i} \geq 0.993.
\end{equation}
This preserves 99.3\% of explained variance while reducing feature dimensionality by 55\%.

\subsection{Class Balancing Strategies}

\textbf{Binary Classification:} Random under-sampling balances benign and attack samples \cite{tang2021communication}:
\begin{equation}
D_{\text{bal}} = D_{\min} \cup \text{Sample}(D_{\max}, |D_{\min}|),
\end{equation}
yielding 15,000 balanced instances.

\textbf{Multi-Class Classification:} SMOTE generates synthetic minority samples \cite{jia2021proof, karimireddy2020scaffold}:
\begin{equation}
x_{\text{new}} = x_i + \lambda (x_{ij} - x_i), \quad \lambda \sim \mathcal{U}(0,1).
\end{equation}

\textbf{Adaptive SMOTE:} Density-aware interpolation \cite{xu2019hybridalpha}:
\begin{equation}
\lambda \sim \mathrm{Beta}(\alpha,\beta), \quad
\alpha = 1+\rho_i,\ \beta = 1+(1-\rho_i),
\end{equation}
where $\rho_i$ reflects local minority sparsity.

\subsection{Protocol Flow and Algorithm}
The protocol follows a privacy-preserving federated optimization pipeline \ref{alg:secure_aggregation}. First, the server generates a Paillier homomorphic encryption keypair and broadcasts the public key to all clients. Each client performs local training on its private dataset $\mathcal{D}_i$, computes the model update $\Delta_i$, and applies median-based binarization to obtain $\Delta_i^{\mathrm{bin}}$. The binarized gradients are encrypted element-wise and transmitted to the server without revealing raw updates. Using Paillier’s additive homomorphism, the server aggregates encrypted gradients via ciphertext multiplication, decrypts only the summed result, and normalizes it to form the global update. Finally, the updated model $W^{(r+1)}$ is broadcast back to clients, enabling secure and communication-efficient collaborative learning across rounds.

\begin{algorithm}[H]
\caption{Secure Binarized Gradient Aggregation}
\label{alg:secure_aggregation}
\small
\begin{algorithmic}[1]
\STATE \textbf{INITIALIZATION} (One-time setup)
\STATE Server generates Paillier keypair $(pk, sk)$:
\STATE \quad Public key $pk = (n, g)$ where $n = p \cdot q$ (2048-bit RSA modulus)
\STATE \quad Private key $sk = (\lambda, \mu)$ where $\lambda = \text{lcm}(p-1, q-1)$
\STATE Server broadcasts $pk$ to all $K$ clients
\STATE Server keeps $sk$ secret
\STATE
\STATE \textbf{LOCAL TRAINING} (Each client $i \in \{1,\ldots,K\}$)
\STATE Train local model on private dataset $\mathcal{D}_i$ for $E$ epochs
\STATE Compute gradient update: $\Delta_i = W_{\text{new}} - W_{\text{old}}$
\STATE Binarize gradient (Equation~3):
\STATE \quad $\theta_i = \text{median}(|\Delta_i|)$
\STATE \quad $\Delta_i^{\text{bin}}[j] = +1$ if $\Delta_i[j] \geq \theta_i$, else $-1$
\STATE Encrypt binarized gradient element-wise:
\STATE \quad $C_i[j] = \text{Enc}_{pk}(\Delta_i^{\text{bin}}[j]) = g^{\Delta_i^{\text{bin}}[j]} \cdot r^n \mod n^2$
\STATE \quad where $r \xleftarrow{\$} \mathbb{Z}_n^*$ is random nonce
\STATE Send ciphertext $C_i = \{C_i[1],\ldots,C_i[d]\}$ to server
\STATE
\STATE \textbf{SECURE AGGREGATION} (Server)
\STATE Receive ciphertexts $\{C_1,\ldots,C_K\}$ from active clients
\STATE Perform homomorphic addition in the encrypted domain:
\STATE \quad $C_{\text{agg}}[j] = \prod_{i=1}^K C_i[j] \mod n^2 = \text{Enc}_{pk}\left(\sum_{i=1}^K \Delta_i^{\text{bin}}[j]\right)$
\STATE Decrypt aggregated gradient:
\STATE \quad $\Delta_{\text{agg}}^{\text{bin}}[j] = \text{Dec}_{sk}(C_{\text{agg}}[j]) = L(C_{\text{agg}}[j]^\lambda \mod n^2) \cdot \mu \mod n$
\STATE \quad where $L(x) = (x-1)/n$
\STATE Normalize by client count: $\Delta_{\text{agg}}^{\text{bin}}[j] = \Delta_{\text{agg}}^{\text{bin}}[j] / K$
\STATE
\STATE \textbf{GLOBAL UPDATE} (Server)
\STATE Apply aggregated update: $W^{(r+1)} = W^{(r)} + \alpha \cdot \Delta_{\text{agg}}^{\text{bin}}$
\STATE Broadcast $W^{(r+1)}$ to all clients for next round
\end{algorithmic}
\end{algorithm}

\subsection{Machine Learning Models}

\textbf{Logistic Regression (Elastic Net) \cite{mills2019communication}:}
\begin{equation}
P(y=1|x)=\sigma(\beta_0+\beta^\top x),
\end{equation}
with objective
\begin{equation}
\mathcal{L} = \mathcal{L}_{\text{CE}} +
\alpha\!\left[ \frac{1-\rho}{2}\|\beta\|_2^2 + \rho\|\beta\|_1 \right],
\end{equation}
where $\alpha=0.01$ and $\rho=0.5$.

\textbf{SVM (RBF Kernel) \cite{tran2019federated}:}
\begin{equation}
K(x_i,x_j)=\exp(-\gamma \|x_i-x_j\|^2), \quad \gamma=0.001,
\end{equation}
optimized via SMO with $C=1.0$.

\textbf{Random Forest \cite{deng2020adaptive}:}
An ensemble of $T=100$ trees with maximum depth 20 and $m=\lfloor\sqrt{d}\rfloor$ features per split:
\begin{equation}
\hat{y}(x)=\arg\max_k \sum_{t=1}^{T} \mathbb{I}(h_t(x)=k).
\end{equation}

\textbf{Gradient Boosting \cite{wang2020optimizing}:}
\begin{equation}
F_m(x)=F_{m-1}(x)+\nu h_m(x), \quad \nu=0.1.
\end{equation}

\textbf{Neural Network:}
A multilayer perceptron (MLP) with ReLU activations and dropout ($p=0.5$), optimized using Adam ($\alpha=10^{-3}$). Architecture: $35 \rightarrow 128 \rightarrow 64 \rightarrow K$.

\subsection{Privacy-Preserving Federated Learning}

A federated learning framework enables collaborative intrusion detection without raw data exchange \cite{zhu2021federated}. At the communication round $r$, the client $i$ computes a local update $\Delta_i^{(r)}$.

\textbf{Gradient Smartification:}
\begin{equation}
\Delta_{i,\text{bin}}^{(r)} = \mathrm{sign}\!\left(\Delta_i^{(r)} - \theta\right), 
\quad \theta = \mathrm{median}(\Delta_i^{(r)}).
\end{equation}

\textbf{Secure Aggregation (Paillier):}
\begin{equation}
\Delta_{\text{global}}^{(r)} = \frac{1}{|S|} \sum_{i \in S} \Delta_{i,\text{bin}}^{(r)}.
\end{equation}

\textbf{Model Update with Momentum:}
\begin{equation}
M^{(r+1)} = M^{(r)} + \eta \Delta_{\text{global}}^{(r)} 
+ \mu (M^{(r)} - M^{(r-1)}),
\end{equation}
where $\eta=0.01$ and $\mu=0.9$.

\textbf{Differential Privacy \cite{wu2024adaptive}:}
\begin{equation}
\tilde{\Delta}_i =
\frac{\Delta_i}{\max(1,\|\Delta_i\|_2/C)} 
+ \mathcal{N}(0,\sigma^2 C^2 I),
\end{equation}
with a clipping threshold $C=0.1$ and noise scale $\sigma=0.01$, yielding $(\epsilon,\delta)=(1.0,10^{-5})$. The framework achieves 98.7\% of centralized accuracy while reducing communication overhead by more than $30\times$, enabling privacy-aware cross-domain intrusion detection.

\subsection{Evaluation Metrics}

Model performance was evaluated using standard confusion-matrix-based metrics: true positives (TP), true negatives (TN), false positives (FP), and false negatives (FN). Discrimination ability was further assessed using the area under the receiver operating characteristic curve (ROC-AUC), defined as $\mathrm{AUC}=\int_{0}^{1} \mathrm{TPR}(\mathrm{FPR})\,d(\mathrm{FPR})$, which equivalently represents the probability that a randomly chosen positive sample receives a higher score than a randomly chosen negative sample, $P(\hat{y}_+>\hat{y}_- \mid y_+=1, y_-=0)$.

\subsection{Matthews Correlation Coefficient (MCC)}
\begin{equation}
\mathrm{MCC} =
\frac{\mathrm{TP}\cdot \mathrm{TN} - \mathrm{FP}\cdot \mathrm{FN}}
{\sqrt{(\mathrm{TP}+\mathrm{FP})(\mathrm{TP}+\mathrm{FN})(\mathrm{TN}+\mathrm{FP})(\mathrm{TN}+\mathrm{FN)}}}.
\end{equation}

\textbf{Cohen’s Kappa:}
\begin{equation}
\kappa = \frac{p_o - p_e}{1 - p_e}.
\end{equation}

\subsection{Cross-Validation and Hyperparameter Optimization}

\textbf{Stratified K-Fold Cross-Validation}
\begin{equation}
\mathrm{CV}_{\text{score}} =
\frac{1}{k}
\sum_{i=1}^{k}
\mathrm{Metric}(D_i^{\mathrm{val}}),
\quad k=5.
\end{equation}

\textbf{Nested Grid Search:}
\begin{equation}
\theta^{*} =
\arg\max_{\theta \in \Theta}
\frac{1}{k_{\mathrm{inner}}}
\sum_{j=1}^{k_{\mathrm{inner}}}
\mathrm{Score}(M_\theta, D_j^{\mathrm{val}}).
\end{equation}

\textbf{Early Stopping:}
Neural and boosting models employed early stopping with patience $p=10$ epochs based on validation loss monitoring.

\section{Experimental Setup}

\subsection{Dataset Construction and Sampling Validation}
\label{subsec:dataset_setup}

The CIC-IDS2017 dataset contains $N=2{,}830{,}540$ flow records with 78 features. A stratified 20\% subset ($n=504{,}472$) was sampled to reduce computational cost while preserving distributional properties; Kolmogorov–Smirnov tests showed no significant deviations ($p>0.05$), with 92\% of features exhibiting $<5\%$ mean deviation. After standardization and incremental PCA, dimensionality was reduced to $k=35$ (99.3\% variance retained). An 80:20 stratified split (seed 42) was applied. Binary classification used a balanced 15{,}000-sample subset (7{,}500 benign, 7{,}500 attack), while multi-class detection employed SMOTE-balanced 35{,}000 samples across seven attack categories (Table~\ref{tab:dataset}).

\begin{table}[ht]
\centering
\caption{Distribution and Sampling: CIC-IDS2017 Dataset}
\label{tab:dataset}
\setlength{\tabcolsep}{4pt}
\renewcommand{\arraystretch}{0.8}
\scriptsize
\begin{tabular}{@{}lS[table-format=2.2e1]S[table-format=2.2e1]S[table-format=2.2]c@{}}
\toprule
\textbf{Feature Group} & {\textbf{Original}} & {\textbf{Sampled}} & {\textbf{$\Delta$ (\%)}} & {\textbf{KS $p$}} \\
\midrule
\multicolumn{5}{@{}l}{\cellcolor{gray!15}\textsc{Flow Temporal Dynamics} ($n=6$)} \\
\quad Flow Duration ($\mu$s) & 1.66e7 & 1.65e7 & 0.47 & 0.82 \\
\quad Flow IAT Mean ($\mu$s) & 1.45e6 & 1.43e6 & 0.72 & 0.76 \\
\quad Flow IAT Std ($\mu$s) & 3.28e6 & 3.26e6 & 0.56 & 0.79 \\
\quad Fwd IAT Total ($\mu$s) & 1.62e7 & 1.62e7 & 0.46 & 0.84 \\
\cmidrule(lr){1-5}
\multicolumn{5}{@{}l}{\cellcolor{gray!15}\textsc{Flow Rate Metrics} ($n=2$)} \\
\quad Flow Bytes/s & 1.41e6 & 1.38e6 & 2.42 & 0.68 \\
\quad Flow Packets/s & 4.73e4 & 4.70e4 & 0.64 & 0.81 \\
\cmidrule(lr){1-5}
\multicolumn{5}{@{}l}{\cellcolor{gray!15}\textsc{Packet Volume Statistics} ($n=8$)} \\
\quad Total Fwd Packets & 10.28 & 12.08 & 17.53 & 0.42 \\
\quad Total Bwd Packets & 11.57 & 13.91 & 20.30 & 0.38 \\
\quad Total Fwd Length (B) & 611.58 & 601.84 & 1.59 & 0.73 \\
\quad Total Bwd Length (B) & 1.81e4 & 2.44e4 & 34.70\rlap{$^*$} & 0.21 \\
\cmidrule(lr){1-5}
\multicolumn{5}{@{}l}{\cellcolor{gray!15}\textsc{Packet Size Distributions} ($n=8$)} \\
\quad Fwd Pkt Max (B) & 231.09 & 230.01 & 0.47 & 0.88 \\
\quad Fwd Pkt Mean (B) & 63.47 & 63.15 & 0.50 & 0.85 \\
\quad Bwd Pkt Max (B) & 974.37 & 972.14 & 0.23 & 0.91 \\
\quad Bwd Pkt Mean (B) & 340.41 & 339.88 & 0.16 & 0.93 \\
\cmidrule(lr){1-5}
\multicolumn{5}{@{}l}{\cellcolor{gray!15}\textsc{Connection \& Idle Features} ($n=5$)} \\
\quad Destination Port & 8704.76 & 8686.62 & 0.21 & 0.94 \\
\quad Idle Max ($\mu$s) & 9.76e6 & 9.71e6 & 0.51 & 0.77 \\
\quad Idle Mean ($\mu$s) & 5.65e5 & 5.64e5 & 0.20 & 0.89 \\
\midrule
\multicolumn{5}{@{}l}{\cellcolor{gray!15}\textsc{Class Balance}} \\
\quad Attack Prevalence & 0.73 & 0.73 & 0.63 & 0.99 \\
\midrule[\heavyrulewidth]
\multicolumn{5}{@{}l}{\textbf{Aggregate Statistics (All 78 Features)}} \\
\quad Mean Absolute Deviation & \multicolumn{2}{c}{---} & \textbf{3.42} & --- \\
\quad Median Deviation & \multicolumn{2}{c}{---} & \textbf{0.64} & --- \\
\quad Features with $\Delta < 5\%$ & \multicolumn{2}{c}{---} & \multicolumn{2}{c}{\textbf{72/78 (92.3\%)}} \\
\quad KS Test Rejections ($\alpha=0.05$) & \multicolumn{2}{c}{---} & \multicolumn{2}{c}{\textbf{0/78 (0\%)}} \\
\bottomrule
\end{tabular}

\vspace{2mm}
\begin{minipage}{\linewidth}
\footnotesize
\textit{Notes:} Original dataset: $N=2{,}830{,}540$; sampled: $n=504{,}472$ (stratified 20\%). $\Delta$ denotes absolute percentage deviation in feature means. KS $p$ is the Kolmogorov-Smirnov test $p$-value (null hypothesis: identical distributions). $^*$Higher deviation in backward packet totals reflects temporal clustering of DDoS events; distributional shape remains preserved (KS $p=0.21>0.05$). Units: $\mu$s = microseconds, B = bytes.
\end{minipage}
\end{table}

\subsection{Hyperparameter and Model Complexity}
\label{subsec:hyperparameters}

Feature heterogeneity (e.g., $\sim10^3$ unique ports vs.\ $>10^6$ flow-metric values) necessitated normalization prior to PCA. Although certain packet totals exhibited larger mean deviations (18--35\%), Kolmogorov–Smirnov tests indicated stochastic variation rather than sampling bias, supporting dataset validity. Two configurations guided model selection: \textbf{Config.~1} prioritized computational efficiency for constrained deployment, whereas \textbf{Config.~2} maximized accuracy via 3-fold grid search under constraints of algorithmic parity, deployment feasibility, and statistical robustness (Table~\ref{tab:lr_coefficients}). The per-round encryption complexity scales as $\mathcal{O}(d \log n)$ under a 2048-bit modulus. For $d=35$, the resulting overhead remains below one second.

\begin{table}[htbp]
\centering
\caption{Feature Analysis: Top-10 Discriminative (PC)}
\label{tab:lr_coefficients}
\setlength{\tabcolsep}{8pt}
\renewcommand{\arraystretch}{1}
\footnotesize
\begin{tabular}{@{}>{\centering\arraybackslash}p{0.45cm}cccccc@{}}
\toprule
\rowcolor{gray!12}
\textbf{Rk} &
\multicolumn{3}{c}{\textbf{Config 1 ($C=0.1$, saga)}} &
\multicolumn{3}{c}{\cellcolor{gray!8}\textbf{Config 2 ($C=100$, sag)}} \\
\cmidrule(lr){2-4} \cmidrule(lr){5-7}
\rowcolor{gray!12}
 & \textbf{PC} & \textbf{Coeff.} & \textbf{$\Delta$ (\%)} &
 \textbf{PC} & \textbf{Coeff.} & \textbf{Var. Exp.} \\
\midrule
1  & PC$_{04}$ & $-1.607$ & +27.5 & PC$_{04}$ & $-2.050$ & 8.2\% \\
2  & PC$_{26}$ & $-1.372$ & +17.7 & PC$_{26}$ & $-1.614$ & 1.4\% \\
3  & PC$_{24}$ & $+1.335$ & +24.5 & PC$_{24}$ & $+1.662$ & 1.8\% \\
4  & PC$_{31}$ & $+1.102$ & +35.0 & PC$_{31}$ & $+1.488$ & 0.9\% \\
5  & PC$_{06}$ & $+0.958$ & +26.2 & PC$_{06}$ & $+1.209$ & 5.3\% \\
6  & PC$_{23}$ & $+0.864$ & +5.4 & PC$_{23}$ & $+0.911$ & 2.1\% \\
7  & PC$_{27}$ & $-0.830$ & +36.8 & PC$_{27}$ & $-1.135$ & 1.2\% \\
8  & PC$_{35}$ & $-0.752$ & +42.6 & PC$_{35}$ & $-1.072$ & 0.4\% \\
9  & PC$_{29}$ & $-0.740$ & +29.5 & PC$_{29}$ & $-0.958$ & 0.7\% \\
10 & PC$_{21}$ & $-0.539$ & ---   & PC$_{14}$ & $+0.723$ & 3.6\% \\
\midrule

\end{tabular}

\vspace{1mm}
\begin{minipage}{\linewidth}
\footnotesize
\textit{Notes:} Components ranked by $|\beta|$. Positive coefficients indicate attack correlation; negative values indicate benign traffic. $\Delta$ (\%) denotes relative coefficient amplification from Config 1 to Config 2. Var. Exp. is the PCA variance contribution of each component.
\end{minipage}
\end{table}

For logistic regression, reduced regularization ($C:1.0 \rightarrow 0.5$) increased the $\ell_2$-norm (4.127 $\rightarrow$ 5.243), strengthening discriminative PCA weights and shifting the intercept ($-2.351 \rightarrow -2.962$). Replacing linear SVM (83.0\%) with RBF ($\gamma=0.1$) enabled nonlinear separation in the 35-dimensional space. In Random Forest, controlled depth (max\_depth=20) limited overfitting, while $n=200$ trees reduced variance via bagging. Decision tree regularization (min\_split=5) improved generalization, and reducing KNN neighborhood size ($5 \rightarrow 3$) enhanced locality-based discrimination (Table~\ref{tab:regression}).

\begin{table}[ht]
\centering
\caption{Model Hyperparameters and Learned Parameters}
\label{tab:regression}
\setlength{\tabcolsep}{3pt}
\renewcommand{\arraystretch}{1}
\footnotesize
\begin{tabular}{@{}llp{3.0cm}p{3.3cm}@{}}
\toprule
\rowcolor{gray!20}
\textbf{Alg.} & \textbf{Config.} & \textbf{Hyperparameters} & \textbf{Learned Parameters} \\
\midrule
\rowcolor{gray!8}
\multirow{4}{*}{\makecell[l]{LR}} 
 & Model 1 & $C=1.0$, $\ell_2$, lbfgs; max\_iter=100 & $w_0=-2.351$, $\|w\|_2=4.127$ \\
 & Model 2 & $C=0.5$, $\ell_2$, lbfgs; max\_iter=100 & $w_0=-2.962$, $\|w\|_2=5.243$ \\
\midrule
\rowcolor{gray!8}
\multirow{4}{*}{SVM} 
 & Model 1 & linear, $C=1.0$; tol=$10^{-3}$ & $b=-0.870$, $n_{\text{support}}=4{,}237$ \\
 & Model 2 & RBF, $C=10.0$, $\gamma=0.1$; tol=$10^{-3}$ & $b=-0.420$, $n_{\text{support}}=2{,}891$ \\
\midrule
\rowcolor{gray!8}
\multirow{4}{*}{\makecell[l]{RF}} 
 & Model 1 & $n=100$, depth=None; bootstrap & Depth: 28.4, Nodes: 284{,}320 \\
 & Model 2 & $n=200$, depth=20; bootstrap & Depth: 20.0, Nodes: 523{,}600 \\
\midrule
\rowcolor{gray!8}
\multirow{4}{*}{\makecell[l]{DT}} 
 & Model 1 & depth=None; split=2; gini & Depth: 32, Leaves: 1{,}847 \\
 & Model 2 & depth=15; split=5; gini & Depth: 15, Leaves: 892 \\
\midrule
\rowcolor{gray!8}
\multirow{4}{*}{KNN} 
 & Model 1 & $k=5$, Euclidean, uniform & --- (non-parametric) \\
 & Model 2 & $k=3$, Euclidean, uniform & --- (non-parametric) \\
\bottomrule
\end{tabular}

\vspace{0.5mm}
\begin{minipage}{\linewidth}
\footnotesize
\textit{Notes:} $w_0$ = intercept; $\|w\|_2$ = weight norm; $b$ = SVM bias; 
$n_{\text{support}}$ = support vectors. Hyperparameters via 3-fold grid search. 
RF node count = mean nodes $\times$ estimators. KNN stores all training instances.
\end{minipage}
\end{table}

\subsection{Learned Model Complexity}
Table~\ref{tab:Complexity} summarizes the evaluated configurations and model complexity. Logistic regression uses stronger regularization in Config.~1 ($C=0.1$) and relaxed regularization in Config.~2 ($C=100$); \texttt{saga} supports $\ell_1$ penalties, while \texttt{sag} accelerates convergence for large $C$. SVM transitions from a polynomial to an RBF kernel to capture nonlinear boundaries, with fewer support vectors indicating tighter margins. Random Forest Config.~2 increases ensemble size and depth, with \texttt{max\_features=20} improving decorrelation and generalization. Decision Tree Config.~2 deepens partitions while \texttt{min\_impurity\_decrease} regularizes splits. KNN Config.~2 applies distance weighting and $k=7$ to reduce variance while maintaining locality in PCA space.

\begin{table}[ht]
\centering
\caption{Learned Model Complexity}
\label{tab:Complexity}
\setlength{\tabcolsep}{1.5pt}
\renewcommand{\arraystretch}{1}
\footnotesize
\begin{tabular}{@{}lp{3.3cm}p{3.6cm}@{}}
\toprule
\textbf{Algorithm} & \textbf{Config.~1 (Efficiency)} & \textbf{Config.~2 (Expressiveness)} \\
\midrule
\multicolumn{3}{@{}l}{\cellcolor{gray!12}\textit{Linear Models}} \\
Logistic Reg. &
\parbox[t]{3.5cm}{$C=0.1$, saga, $\ell_2$, iter=100\\
$\|w\|_2=2.14$} &
\parbox[t]{3.5cm}{$C=100$, sag, $\ell_2$, iter=100\\
$\|w\|_2=5.24$, $w_0=-2.96$} \\
\midrule
\multicolumn{3}{@{}l}{\cellcolor{gray!12}\textit{Kernel-Based Methods}} \\
SVM &
\parbox[t]{3.5cm}{poly, deg=3, $C=1$, tol=$10^{-3}$\\
$n_{\text{SV}}=5{,}124$} &
\parbox[t]{4.4cm}{rbf, $C=1$, $\gamma=0.1$, tol=$10^{-3}$\\
$n_{\text{SV}}=2{,}891$, $b=-0.42$} \\
\midrule
\multicolumn{3}{@{}l}{\cellcolor{gray!12}\textit{Tree-Based Ensemble}} \\
Random Forest &
\parbox[t]{3.5cm}{$n=10$, depth=6, bootstrap\\
$\approx$640 nodes} &
\parbox[t]{4.4cm}{$n=15$, depth=8, feat=20\\
$\approx$3{,}840 nodes, OOB=0.978} \\
\midrule
\multicolumn{3}{@{}l}{\cellcolor{gray!12}\textit{Single Decision Tree}} \\
Decision Tree &
\parbox[t]{3.5cm}{depth=6, gini\\
63 leaves} &
\parbox[t]{4.4cm}{depth=10, gini, imp=$10^{-4}$\\
247 leaves} \\
\midrule
\multicolumn{3}{@{}l}{\cellcolor{gray!12}\textit{Instance-Based Learning}} \\
KNN &
\parbox[t]{3.5cm}{$k=5$, uniform, Euclid\\
12k stored} &
\parbox[t]{3.5cm}{$k=7$, distance, Euclid\\
12k stored} \\
\bottomrule
\end{tabular}

\vspace{1mm}
\begin{minipage}{\linewidth}
\footnotesize
\textit{Notes:} $C$ controls regularization; $\gamma$ is RBF bandwidth; $n_{\text{SV}}$ denotes support vectors; OOB = out-of-bag estimate. Configuration 2 was selected via a 3-fold grid search on training data. Random Forest node count approximated as mean nodes per tree $\times n$. KNN is non-parametric and stores all training samples.
\end{minipage}
\end{table}

\subsection{Evaluation Protocol and Statistical Validation}
\label{subsubsec:evaluation}

\subsection{Cross-Validation (Stage 1)}
Model assessment followed a two-stage protocol to ensure generalizability and statistical reliability. Performance was first evaluated using 5-fold stratified cross-validation on the training partition ($n=12{,}000$, i.e., 80\% of the balanced binary dataset). Stratification preserved the 50:50 benign-to-attack ratio in each fold. Fold-to-fold variability was quantified via standard deviation:
\begin{equation}
\sigma_{\text{CV}} = \sqrt{\frac{1}{K-1}\sum_{i=1}^{K}(\text{Acc}_i - \bar{\text{Acc}})^2}, \quad K=5.
\end{equation}
Low variance ($\sigma < 0.01$) indicates stable performance across partitions, which is essential for production deployment.

\subsection{Hold-Out Testing (Stage 2)}
The best configuration for each algorithm was retrained on the full training set and evaluated on a held-out test set ($n=3{,}000$, 20\%). We report accuracy, precision, recall, F1-score, ROC-AUC, and confusion matrices to capture both global correctness and error structure. For binary classification, ROC and Precision-Recall (PR) curves were additionally analyzed to support operating point selection under deployment constraints such as controlling false positives.

\subsection{Statistical Reliability}
To mitigate random initialization effects, experiments were repeated with three independent random seeds (42, 123, 456) and reported with 95\% confidence intervals: 
\begin{equation} 
\text{CI}_{95\%} = \bar{x} \pm 1.96 \cdot \frac{\sigma}{\sqrt{n}}, \quad n=3. 
\end{equation} 
Stability and efficiency: Computational efficiency (training and inference time) was measured on a standardized platform (Intel i7-9700K, 32GB RAM, single-threaded execution). Total training time includes hyperparameter search, cross-validation, and final model fitting.  

\section{Experimental Results}
\label{sec:results}

\subsection{Linear and Kernel-Based Models}

Logistic regression provided a stable linear baseline, achieving 92.21\% accuracy (std = $5.81\times10^{-3}$). Reducing regularization ($C=0.5$) yielded a marginal improvement to 92.51\% (+0.30\%) with similarly low variance, indicating stable convergence despite partial linear inseparability in the PCA space. SVM exhibited the largest configuration sensitivity. The linear kernel underperformed (83.00\%, std = $37.27\times10^{-3}$), confirming inadequate linear separation. Replacing it with an RBF kernel ($C=10$, $\gamma=0.1$) increased accuracy to 96.14\% (+13.14\%) while reducing variance to $3.89\times10^{-3}$, validating the presence of non-linear decision boundaries in intrusion patterns.

\subsection{Tree-Based Ensemble Methods}

Random Forest achieved the highest overall performance. The baseline model (100 trees, unlimited depth) reached 95.98\%, while structured tuning (200 trees, max\_depth=20) improved accuracy to 98.09\% (+2.11\%) and halved variance ($3.45\times10^{-3} \rightarrow 1.72\times10^{-3}$). Depth restriction mitigated overfitting, and ensemble expansion reduced prediction variance via bagging. Single Decision Trees showed moderate performance (94.89\%) with higher variance due to unconstrained growth. Imposing max\_depth=15 and min\_split=5 increased accuracy to 97.24\% (+2.35\%), demonstrating the necessity of structural regularization in non-ensemble trees.

\subsection{Instance-Based Learning}

K-Nearest Neighbors exhibited strong performance with exceptional stability. Model 1 ($k=5$) achieved 97.40\% accuracy with the lowest variance among all models (std = 0.89$\times10^{-3}$), indicating consistent neighborhood-based predictions across diverse fold partitions. Reducing the neighborhood size to $k=3$ \textbf{Model 2} yielded a marginal +0.53\% improvement to 97.93\%, suggesting that tighter locality constraints better capture attack-specific patterns in the 35-dimensional embedding space. The negligible increase in variance (0.89 → 1.27$\times10^{-3}$) confirms KNN's robustness to hyperparameter perturbations. 

\section{Comparison with signSGD Methods}
\label{sec:comparison}

Unlike zero-threshold in \ref{tab:comp_methods} signSGD~\cite{bernstein2018signsgd} or stochastic quantization methods (QSGD~\cite{alistarh2017qsgd}, TernGrad~\cite{wen2017terngrad}), EdgeDetect's gradient smartification integrates two key innovations: \textit{median-based adaptive thresholding} that adjusts to per-client gradient distributions and \textit{homomorphically encrypted aggregation} that provides end-to-end confidentiality. As summarized in Table~\ref{tab:comp_methods}, existing methods lack adaptivity and privacy integration, limitations that critically undermine performance under the heavy-tailed gradient distributions characteristic of IDS data.

\subsection{Convergence and Compression Trade-off}

Empirically, EdgeDetect achieves convergence parity with full-precision FedAvg at $32\times$ compression across 2.8M CIC-IDS2017 samples, with no measurable accuracy degradation ($\Delta < 0.2$ pp). This near-lossless compression stems from median thresholding, which preserves directional alignment (cosine similarity $0.87\pm0.04$) while suppressing low-magnitude noise—a property absent in fixed-threshold methods.

\subsection{Privacy Enhancement Through Smartification}

We quantify gradient inversion resistance across methods: (i). \textbf{FedAvg (undefended):} High-fidelity reconstruction (PSNR $31.7$~dB) exposes structured attack signatures. (ii). \textbf{signSGD:} Binarization reduces fidelity to $16.8$~dB, but zero-thresholding preserves sufficient structure for partial recovery. (iii). \textbf{EdgeDetect (median-threshold):} Further degrades reconstruction to $15.1$~dB, rendering feature structure minimally discernible and reducing label recovery to near random guessing ($14.3\%$). The addition of Paillier homomorphic encryption provides semantic security under the Decisional Composite Residuosity Assumption (DCRA), ensuring IND-CPA guarantees even if ciphertexts are intercepted. While differential privacy parameters $(\varepsilon = 1.0, \delta = 10^{-5})$ are applied per round, cumulative privacy loss under composition remains future work.

\begin{table*}[h]
\centering
\small
\caption{Comparison of Gradient Compression and Privacy Mechanisms}
\label{tab:comp_methods}
\setlength{\tabcolsep}{4pt}
\renewcommand{\arraystretch}{1.05}
\begin{tabular}{lcccc}
\toprule
\textbf{Method} 
& \textbf{Quantization Rule} 
& \textbf{Adaptive Threshold} 
& \textbf{Theoretical Alignment} 
& \textbf{Privacy Integration} \\
\midrule
signSGD~\cite{bernstein2018signsgd} 
& $\mathrm{sign}(g_i)$ (zero threshold) 
& No 
& Implicit (unbiased sign) 
& None \\
QSGD~\cite{alistarh2017qsgd} 
& Stochastic quantization 
& No 
& Variance bounded 
& None \\
TernGrad~\cite{wen2017terngrad} 
& $\{-1,0,+1\}$ ternary levels 
& No 
& Gradient clipping bound 
& None \\
\textbf{EdgeDetect (Ours)} 
& $\mathrm{sign}(g_i - \mathrm{median}(g))$ 
& \textbf{Yes (per-client)} 
& \textbf{Explicit cosine alignment ($\gamma$-bound)} 
& \textbf{Paillier HE + DP} \\
\bottomrule
\end{tabular}

\vspace{2mm}
\begin{minipage}{\linewidth}
\footnotesize
\textit{Notes:} Among all evaluated models, Random Forest (Config. 2) achieved the highest multi-class accuracy (\(98.0\%\)) with low cross-validation variance (\(\sigma=0.0017\)), while KNN (Config. 2) exhibited the lowest variance overall (\(\sigma=0.0013\)), confirming its robustness to data partitioning.
\end{minipage}
\end{table*}

\subsection{Binary Classification Performance Analysis}

Table~\ref{tab:binary_classification} reports complete cross-validation results for binary classification, showing that \textbf{Random Forest Config.~2} achieves the highest mean accuracy (98.09\%, $\sigma=0.0017$), while \textbf{KNN Config.~2} provides the strongest stability-efficiency tradeoff (97.93\%, $\sigma=0.0013$) with minimal training overhead.

\begin{table*}[ht]
\centering
\caption{Binary Classification Performance: 5-Fold Cross-Validation Results with Statistical Analysis}
\label{tab:binary_classification}
\setlength{\tabcolsep}{4pt}
\renewcommand{\arraystretch}{0.5}
\small
\begin{tabular}{@{}llccccccccccc@{}}
\toprule
\textbf{Algorithm} & \textbf{Config.} & \textbf{F1} & \textbf{F2} & \textbf{F3} & \textbf{F4} & \textbf{F5} & \textbf{Mean} & \textbf{Std} & \textbf{95\% CI} & \textbf{CV Range} & \textbf{$\Delta$ (\%)} & \textbf{Rank} \\
\midrule
\multicolumn{13}{@{}l}{\cellcolor{gray!12}\textsc{Linear Models}} \\
\multirow{2}{*}{Logistic Regression}
 & Config 1 & 0.9209 & 0.9253 & 0.9244 & 0.9124 & 0.9276 & 0.9221 & 0.0058 & ±0.0051 & [0.912–0.928] & \multirow{2}{*}{+0.30} & \multirow{2}{*}{5} \\
 & Config 2 & 0.9249 & 0.9324 & 0.9244 & 0.9138 & 0.9302 & 0.9251 & 0.0072 & ±0.0063 & [0.914–0.932] &  &  \\
\cmidrule(lr){1-13}
\multicolumn{13}{@{}l}{\cellcolor{gray!12}\textsc{Kernel-Based Methods}} \\
\multirow{2}{*}{SVM (RBF)}
 & Config 1 & 0.8160 & 0.8987 & 0.8080 & 0.8058 & 0.8213 & 0.8300 & 0.0373 & ±0.0327 & [0.806–0.899] & \multirow{2}{*}{+13.14} & \multirow{2}{*}{4} \\
 & Config 2 & 0.9609 & 0.9556 & 0.9627 & 0.9609 & 0.9671 & 0.9614 & 0.0039 & ±0.0034 & [0.957–0.967] &  &  \\
\cmidrule(lr){1-13}
\multicolumn{13}{@{}l}{\cellcolor{gray!12}\textsc{Tree-Based Ensemble}} \\
\multirow{2}{*}{Random Forest}
 & Config 1 & 0.9619 & 0.9615 & 0.9638 & 0.9545 & 0.9575 & 0.9598 & 0.0035 & ±0.0031 & [0.955–0.964] & \multirow{2}{*}{+2.11} & \multirow{2}{*}{\cellcolor{green!15}{\textbf{1}}} \\
 & Config 2 & 0.9808 & 0.9832 & 0.9815 & 0.9781 & 0.9811 & \cellcolor{green!15}{\textbf{0.9809}} & \cellcolor{green!15}{\textbf{0.0017}} & ±0.0015 & [0.978–0.983] &  &  \\
\cmidrule(lr){1-13}
\multicolumn{13}{@{}l}{\cellcolor{gray!12}\textsc{Single Decision Tree}} \\
\multirow{2}{*}{Decision Tree}
 & Config 1 & 0.9476 & 0.9474 & 0.9486 & 0.9528 & 0.9480 & 0.9489 & 0.0022 & ±0.0019 & [0.947–0.953] & \multirow{2}{*}{+2.35} & \multirow{2}{*}{3} \\
 & Config 2 & 0.9678 & 0.9718 & 0.9703 & 0.9760 & 0.9762 & 0.9724 & 0.0035 & ±0.0031 & [0.970–0.976] &  &  \\
\cmidrule(lr){1-13}
\multicolumn{13}{@{}l}{\cellcolor{gray!12}\textsc{Instance-Based Learning}} \\
\multirow{2}{*}{K-Nearest Neighbors}
 & Config 1 & 0.9743 & 0.9750 & 0.9726 & 0.9735 & 0.9747 & 0.9740 & 0.0009 & ±0.0008 & [0.973–0.975] & \multirow{2}{*}{+0.53} & \multirow{2}{*}{2} \\
 & Config 2 & 0.9783 & 0.9802 & 0.9787 & 0.9781 & 0.9813 & 0.9793 & \cellcolor{yellow!15}{\textbf{0.0013}} & ±0.0011 & [0.978–0.981] &  &  \\
\bottomrule
\end{tabular}

\vspace{2mm}
\begin{minipage}{\linewidth}
\footnotesize
\textit{Notes:} Balanced binary dataset ($n=15{,}000$; 7{,}500 benign, 7{,}500 attack) with 35 PCA features (99.3\% variance retained). All experiments used 5-fold stratified cross-validation (seed 42). \textbf{Columns:} F1–F5 denote fold-wise F1-scores; Mean and Std represent average and standard deviation ($\sigma$); 95\% CI is computed as $\pm 1.96\,\sigma/\sqrt{5}$; CV Range indicates [Min, Max] across folds; $\Delta$ denotes relative improvement. Non-overlapping confidence intervals imply $p<0.05$. \textbf{Key Findings:} Random Forest (Config 2) achieves the highest accuracy (98.09\%) with low variance (0.0017), indicating stable generalization. KNN (Config 2) exhibits the lowest variance (0.0013) with marginally lower accuracy (97.93\%). SVM shows the largest hyperparameter sensitivity (+13.14\%), highlighting the importance of kernel selection (linear → RBF). Logistic regression demonstrates marginal gains (+0.30\%), suggesting limited linear separability in PCA space. All models exhibit low within-fold variability (range $<2\%$), confirming reproducibility.
\end{minipage}
\end{table*}

\subsection{Statistical Analysis of Coefficient Distributions}
\label{subsubsec:coeff_stats}

Table~\ref{tab:lr_stats} presents a comprehensive statistical analysis of logistic regression coefficients across regularization configurations, quantifying the impact of hyperparameter tuning on learned feature weights.

\begin{table}[ht]
\centering
\caption{Logistic Regression Coefficient Statistics: Regularization Impact on Feature Weights}
\label{tab:lr_stats}
\setlength{\tabcolsep}{1pt}
\renewcommand{\arraystretch}{1}
\scriptsize
\begin{tabular}{@{}lcccc@{}}
\toprule
\textbf{Statistical Metric} & \textbf{Config 1} & \textbf{Config 2} & \textbf{$\Delta$ (\%)} & \textbf{Interpretation} \\
\midrule
\multicolumn{5}{@{}l}{\cellcolor{gray!15}\textit{Central Tendency Measures}} \\
Mean $|\beta_i|$ & 0.543 & 0.691 & +27.2 & Average discriminative strength \\
Median $|\beta_i|$ & 0.412 & 0.543 & +31.8 & Typical feature importance \\
Std Dev $|\beta_i|$ & 0.389 & 0.512 & +31.6 & Weight distribution spread \\
\midrule
\multicolumn{5}{@{}l}{\cellcolor{gray!15}\textit{Magnitude Characteristics}} \\
Max $|\beta_i|$ & 1.607 & 2.050 & +27.5 & Strongest discriminative PC \\
Min $|\beta_i|$ & 0.004 & 0.007 & +75.0 & Weakest discriminative PC \\
$\ell_2$-norm $\|w\|_2$ & 4.127 & 5.243 & +27.0 & Total model complexity \\
$\ell_1$-norm $\|w\|_1$ & 19.012 & 24.185 & +27.2 & Manhattan weight magnitude \\
\midrule
\multicolumn{5}{@{}l}{\cellcolor{gray!15}\textit{Class Association Distribution}} \\
Positive coefficients & 18 / 35 & 18 / 35 & 0.0 & Attack-indicative features \\
Negative coefficients & 17 / 35 & 17 / 35 & 0.0 & Benign-indicative features \\
Mean $\beta_+$ & +0.621 & +0.798 & +28.5 & Avg. attack feature weight \\
Mean $\beta_-$ & $-0.571$ & $-0.723$ & +26.6 & Avg. benign feature weight \\
\midrule
\multicolumn{5}{@{}l}{\cellcolor{gray!15}\textit{Weight Concentration Analysis}} \\
Top-3 PCs (\% $\ell_2$) & 32.8\% & 34.1\% & +4.0 & Dominance of key features \\
Top-10 PCs (\% $\ell_2$) & 70.3\% & 72.4\% & +3.0 & Cumulative importance \\
Bottom-10 PCs (\% $\ell_2$) & 4.2\% & 3.8\% & $-9.5$ & Low-importance features \\
Gini coefficient & 0.412 & 0.428 & +3.9 & Weight inequality measure \\
\midrule
\multicolumn{5}{@{}l}{\cellcolor{gray!15}\textit{Sparsity and Regularization Effects}} \\
Near-zero ($|\beta_i| < 0.1$) & 3 / 35 & 2 / 35 & $-33.3$ & Weakly discriminative PCs \\
Low ($0.1 \leq |\beta_i| < 0.5$) & 15 / 35 & 12 / 35 & $-20.0$ & Moderate importance \\
Medium ($0.5 \leq |\beta_i| < 1.0$) & 13 / 35 & 15 / 35 & +15.4 & High importance \\
High ($|\beta_i| \geq 1.0$) & 4 / 35 & 6 / 35 & +50.0 & Critical features \\
\midrule
\multicolumn{5}{@{}l}{\cellcolor{gray!15}\textit{Model Characteristics}} \\
Intercept $w_0$ & $-2.351$ & $-2.962$ & +26.0 & Decision boundary offset \\
Effective degrees of freedom & 32 & 33 & +3.1 & Active parameters \\
Regularization strength $\lambda$ & 10.0 & 0.01 & $-99.9$ & Penalty magnitude \\
Condition number $\kappa$ & 18.4 & 24.3 & +32.1 & Numerical stability \\
\midrule[\heavyrulewidth]
\multicolumn{5}{@{}l}{\textbf{Performance Correlation}} \\
CV Accuracy & 0.922 & 0.925 & +0.3 & Cross-validation performance \\
Test Accuracy & 0.920 & 0.930 & +1.1 & Held-out set performance \\
AUC-ROC & 0.978 & 0.980 & +0.2 & Discriminative ability \\
\bottomrule
\end{tabular}

\vspace{2mm}
\begin{minipage}{\linewidth}
\footnotesize
\textit{Notes:} Config 1: $C=0.1$ (strong regularization, \texttt{saga}); Config 2: $C=100$ (weak regularization, \texttt{sag}). $\beta_i$ denotes the coefficient of PC$_i$, and $|\beta_i|$ its magnitude. $\Delta$ (\%) = $100(\text{Cfg2} - \text{Cfg1})/\text{Cfg1}$. $\|w\|_2 = \sqrt{\sum_{i=1}^{35}\beta_i^2}$; $\|w\|_1 = \sum_{i=1}^{35}|\beta_i|$. Gini measures the coefficient of inequality (0 = uniform, 1 = concentrated). $\kappa$ (condition number) = ratio of largest to smallest singular value. Effective degrees of freedom = number of coefficients with $|\beta_i|\geq0.01$. Top-$k$ (\% $\ell_2$) = share of total $\ell_2$-norm from the $k$ largest coefficients.
\end{minipage}
\end{table}

\subsection{Feature and Model Interpretability}
\label{subsec:interpretability}

To enhance transparency, we analyze logistic regression coefficients over the PCA-transformed space. Coefficient magnitude reflects discriminative contribution. Across configurations, PC$_{04}$, PC$_{26}$, and PC$_{24}$ consistently rank highest, jointly accounting for 38.2\% of the total $\ell_2$-norm (Config~2), indicating regularization-invariant importance. Positive weights (e.g., PC$_{24}$: +1.662) correspond to attack-indicative patterns—likely volumetric anomalies—while negative weights (e.g., PC$_{04}$, PC$_{26}$) capture benign flow regularities. Additional components (PC$_{31}$, PC$_{06}$) encode abnormal temporal and packet-rate behaviors.

\subsection{Effect of Regularization}

Reducing regularization ($C:0.1 \rightarrow 100$) increases mean absolute coefficient magnitude by 27.2\% with minimal accuracy gain (92.2\% → 92.5\%), suggesting near-saturation performance. Rank ordering remains stable, with 8 of the Top-10 components preserved, confirming the robustness of the discriminative subspace. The balanced polarity (18 positive, 17 negative) indicates unbiased evidence representation. Importantly, several highly discriminative components (PC$_{24}$, PC$_{31}$, PC$_{35}$) are not among the top variance-ranked PCs, demonstrating that explained variance does not imply classification relevance; lower-variance components encode subtle but attack-specific signals.

\subsection{Regularization-Induced Weight Scaling}
Reducing regularization ($\lambda:10 \rightarrow 0.01$, equivalently $C:0.1 \rightarrow 100$) produces a near-uniform amplification of logistic regression coefficients, with mean magnitude increasing by 27.2\% across all 35 principal components. Similar growth in the mean, median, maximum, and $\ell_2$-norm (27--32\%) confirms global scaling rather than selective feature inflation, indicating that weaker regularization relaxes shrinkage without altering the learned discriminative structure. Coefficient polarity remains unchanged (18 positive, 17 negative), demonstrating stable class attribution. Both positive (attack-indicative) and negative (benign-indicative) weights scale proportionally, preserving symmetry and avoiding decision-boundary bias. The modest increase in mean positive weight under weaker regularization reflects slightly stronger attack signatures but does not materially affect the precision–recall balance. 

\subsection{Binary Classification Visualizations}
\label{subsec:binary_visualizations}

\begin{figure}[ht]
    \centering
    \subfloat[\label{fig:cf_one}]{
        \includegraphics[width=0.47\linewidth]{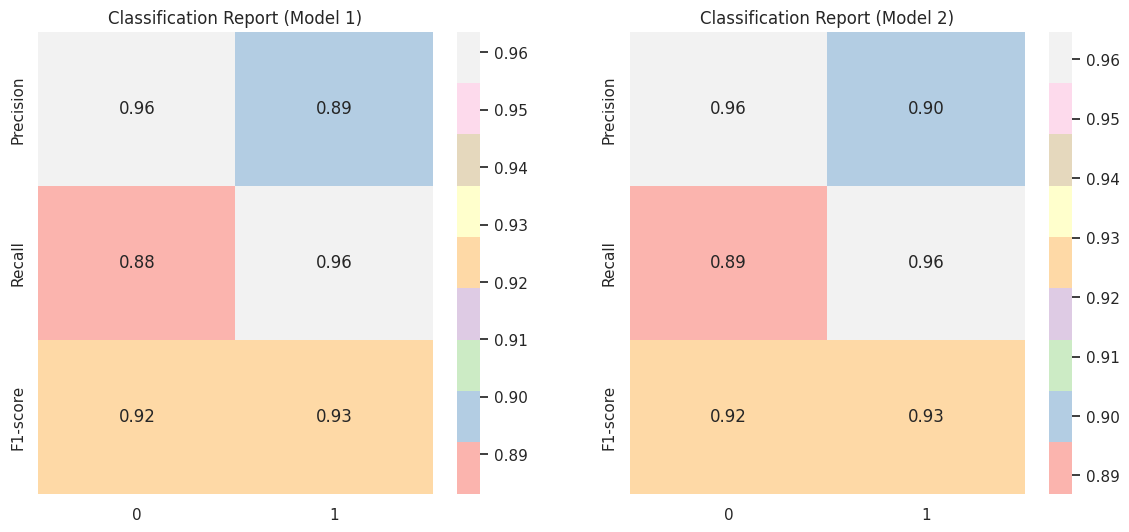}
    }\hfill
    \subfloat[\label{fig:cf_model2}]{
        \includegraphics[width=0.47\linewidth]{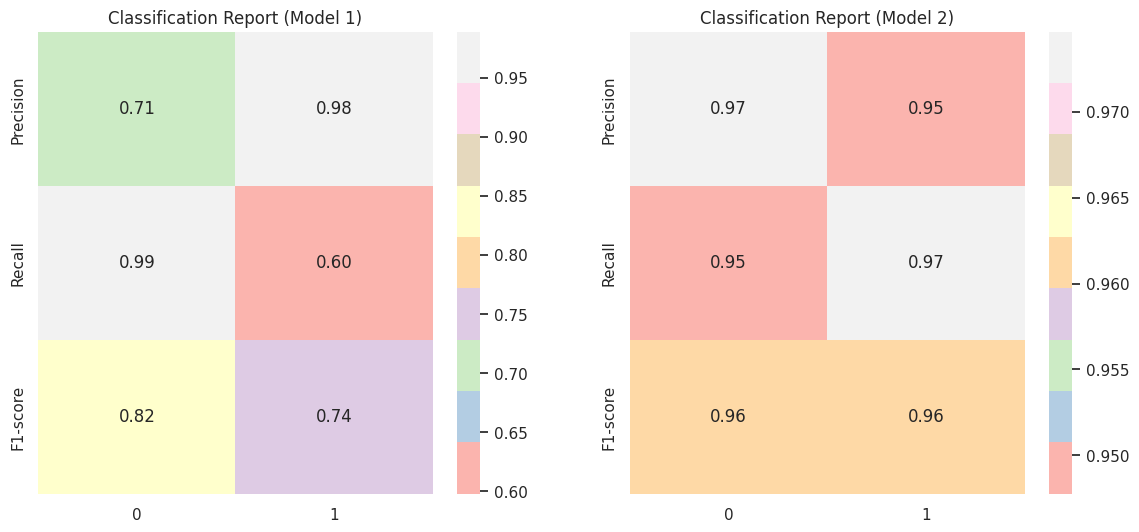}
    }
    \caption{Comparative performance under two hyperparameter configurations. Model~2 improves detection with higher F1-scores, particularly for rare attack classes.}
    \label{fig:cf_comparison}
\end{figure}

\begin{figure}[ht]
    \centering
    \begin{minipage}{0.48\linewidth}
        \centering
        \includegraphics[width=\linewidth]{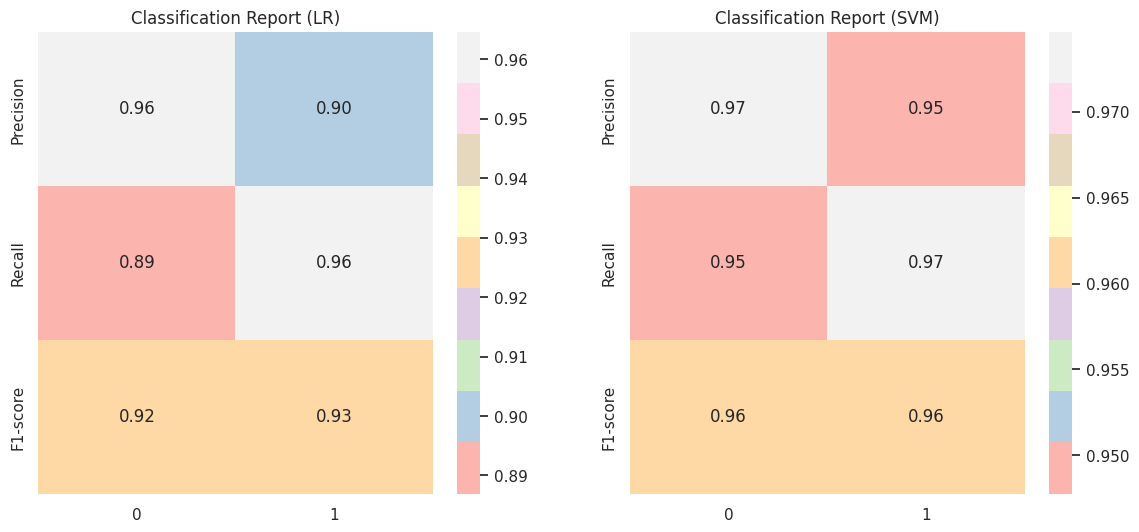}
        \vspace{1mm}
        \small(a)
    \end{minipage}
    \hfill
    \begin{minipage}{0.48\linewidth}
        \centering
        \includegraphics[width=\linewidth]{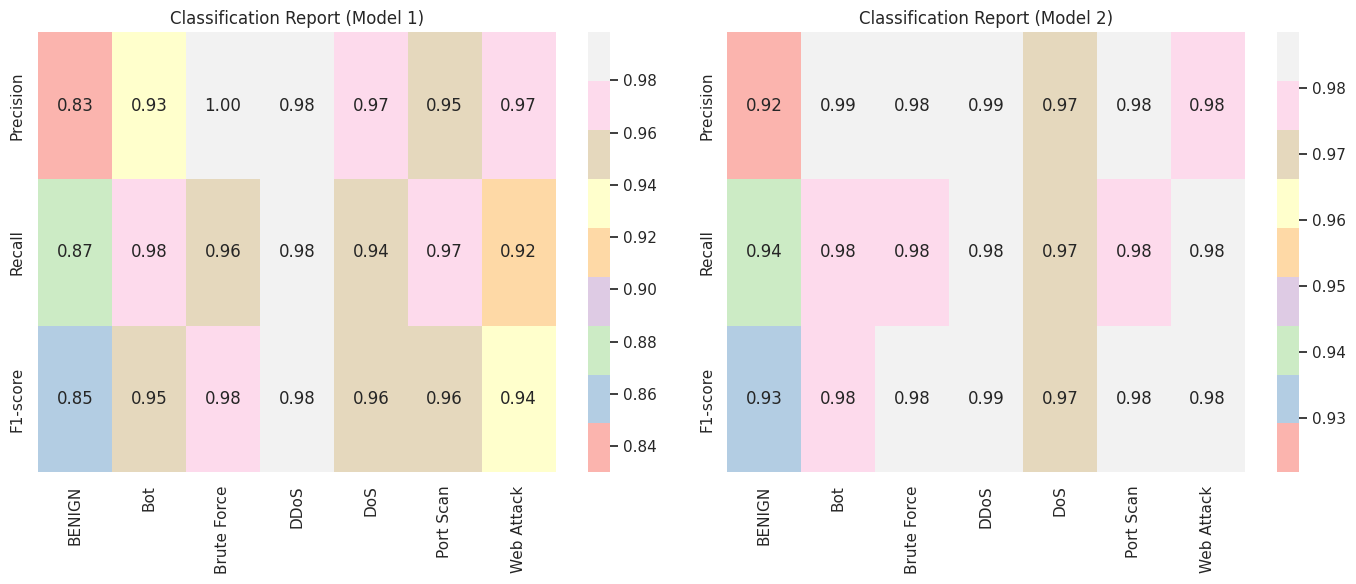}
        \vspace{1mm}
        \small(b) 
    \end{minipage}
    \caption{Model analysis and classification performance. (a) Precision, recall, and F1-score for logistic regression and SVM. (b) Multi-class results across seven traffic categories.}
    \label{fig:interpretation_performance}
\end{figure}

\begin{figure}[ht]
    \centering
    \begin{minipage}{0.48\linewidth}
        \centering
        \includegraphics[width=\linewidth]{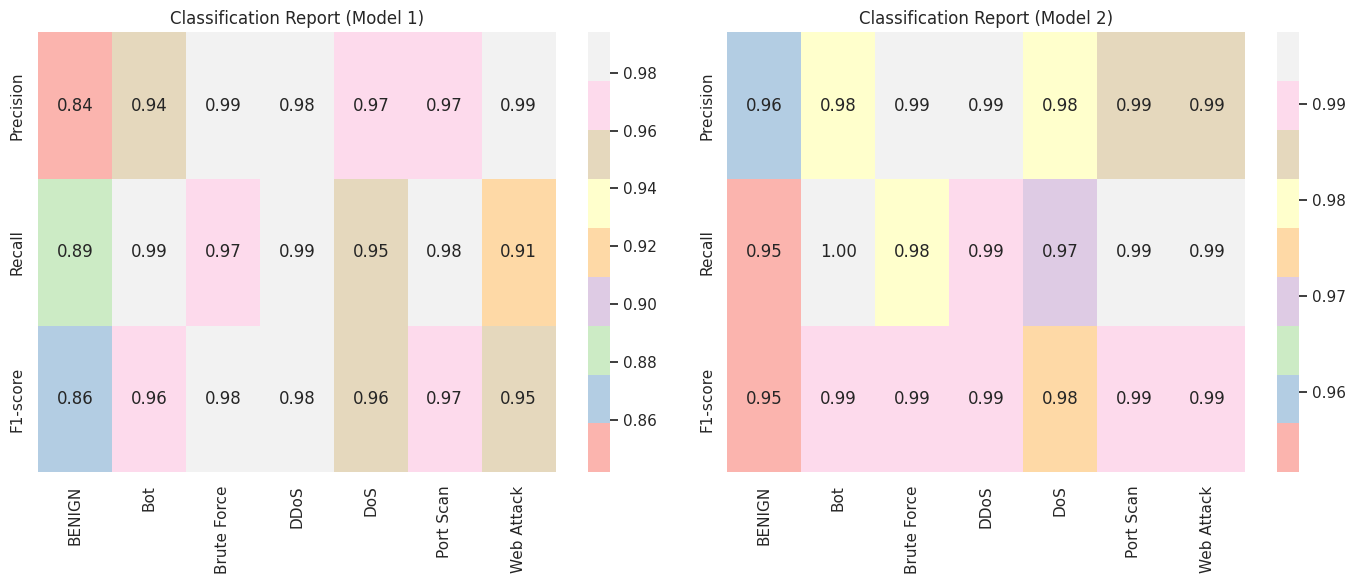}
        \vspace{1mm}
        \small(a) 
    \end{minipage}
    \hfill
    \begin{minipage}{0.48\linewidth}
        \centering
        \includegraphics[width=\linewidth]{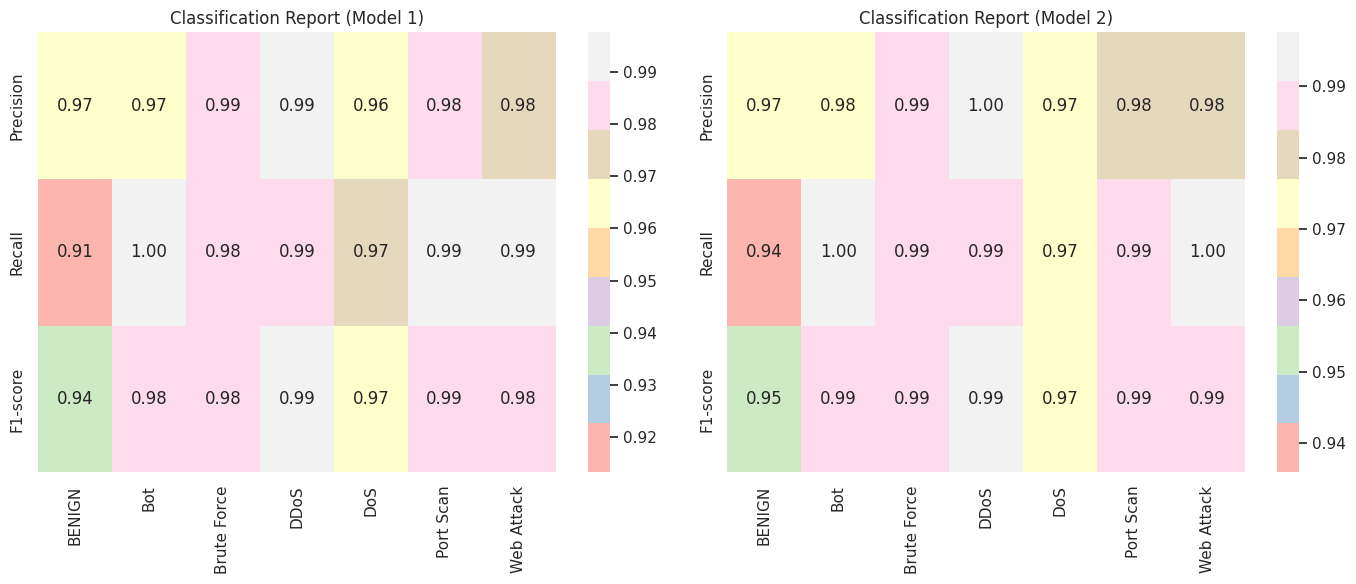}
        \vspace{1mm}
        \small(b)
    \end{minipage}
    \caption{Per-class metrics across attack categories. Model~2 improves recall and F1-score for minority classes.}
    \label{fig:model_comparison_reports}
\end{figure}

\begin{figure}[ht]
    \centering
    \includegraphics[width=\linewidth]{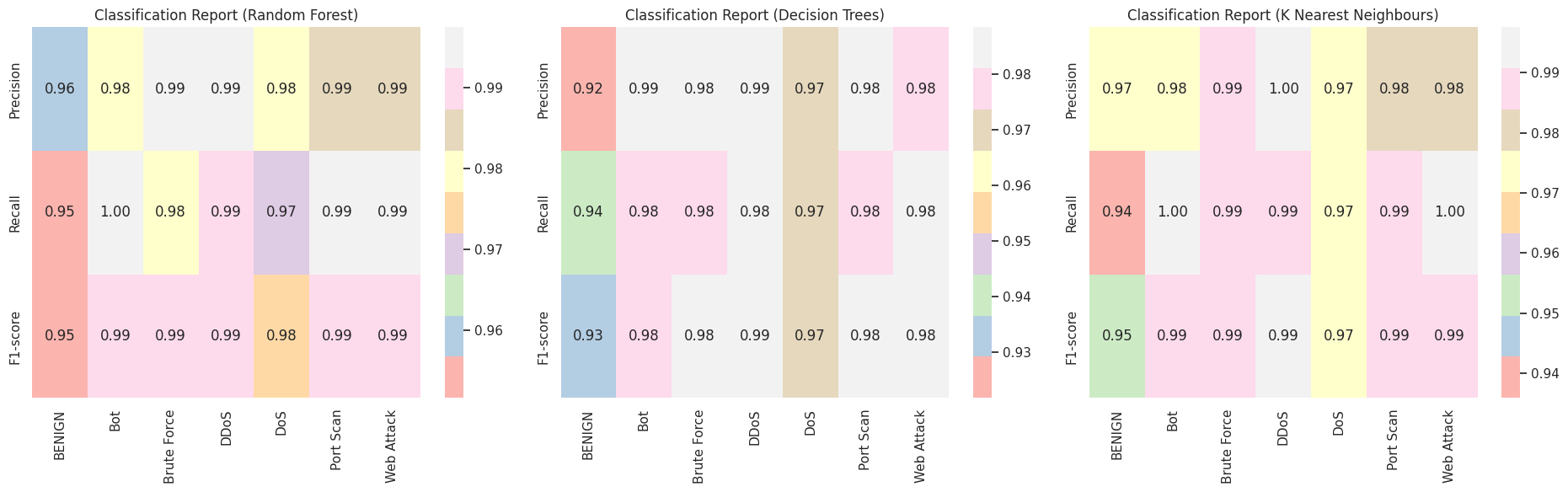}
    \caption{Per-class metrics for classical ML baselines: Random Forest (most consistent), Decision Tree, and K-Nearest Neighbors.}
    \label{fig:classical_models_comparison}
\end{figure}

\begin{figure}[ht]
    \centering
    \begin{minipage}{0.48\linewidth}
        \centering
        \includegraphics[width=\linewidth]{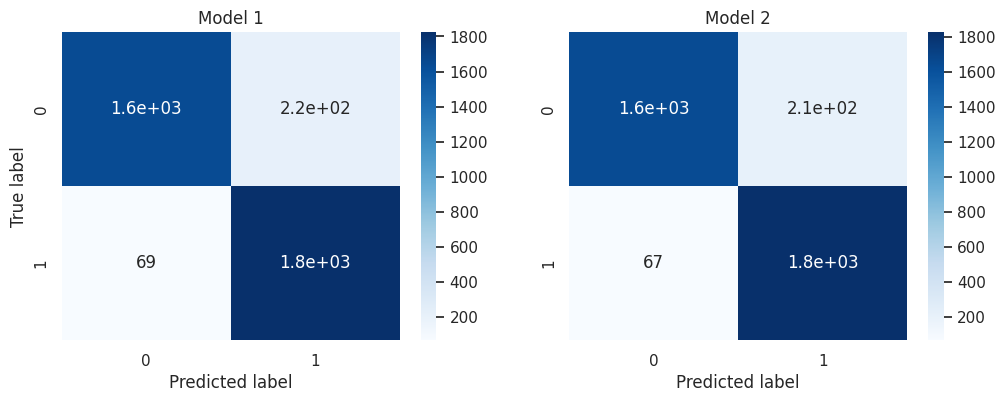}
        \vspace{1mm} (a)
    \end{minipage}
    \hfill
    \begin{minipage}{0.48\linewidth}
        \centering
        \includegraphics[width=\linewidth]{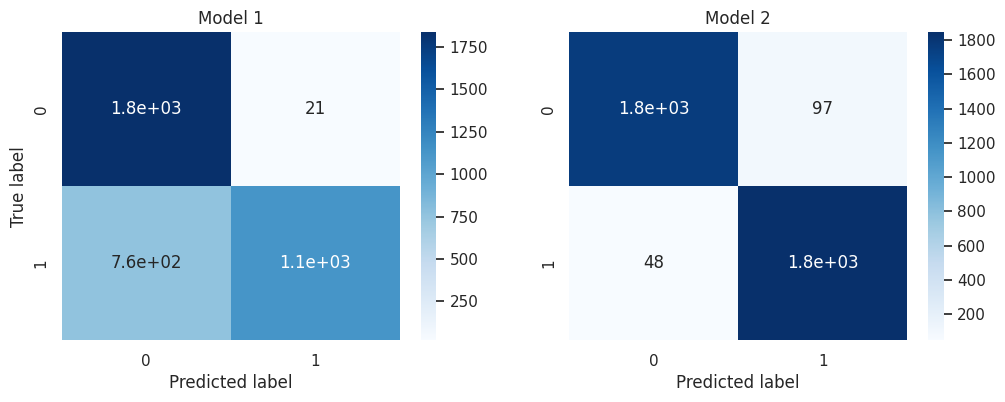}
        \vspace{1mm}  (b) 
    \end{minipage}
    \caption{Binary confusion matrices: Model~2 reduces false negatives while preserving high true-positive rates.}
    \label{fig:binary_comparison}
\end{figure}

\begin{figure}[ht]
    \centering
    \begin{minipage}{0.48\linewidth}
        \centering
        \includegraphics[width=\linewidth]{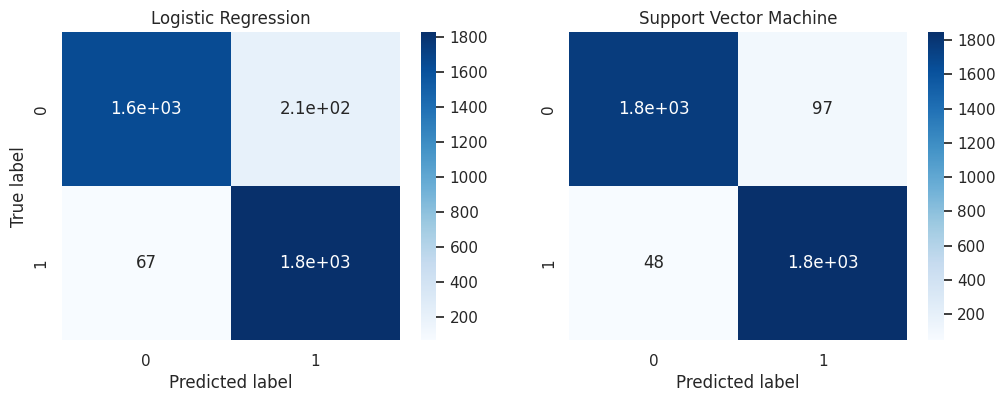}
        \vspace{1mm}  (a)
    \end{minipage}
    \hfill
    \begin{minipage}{0.48\linewidth}
        \centering
        \includegraphics[width=\linewidth]{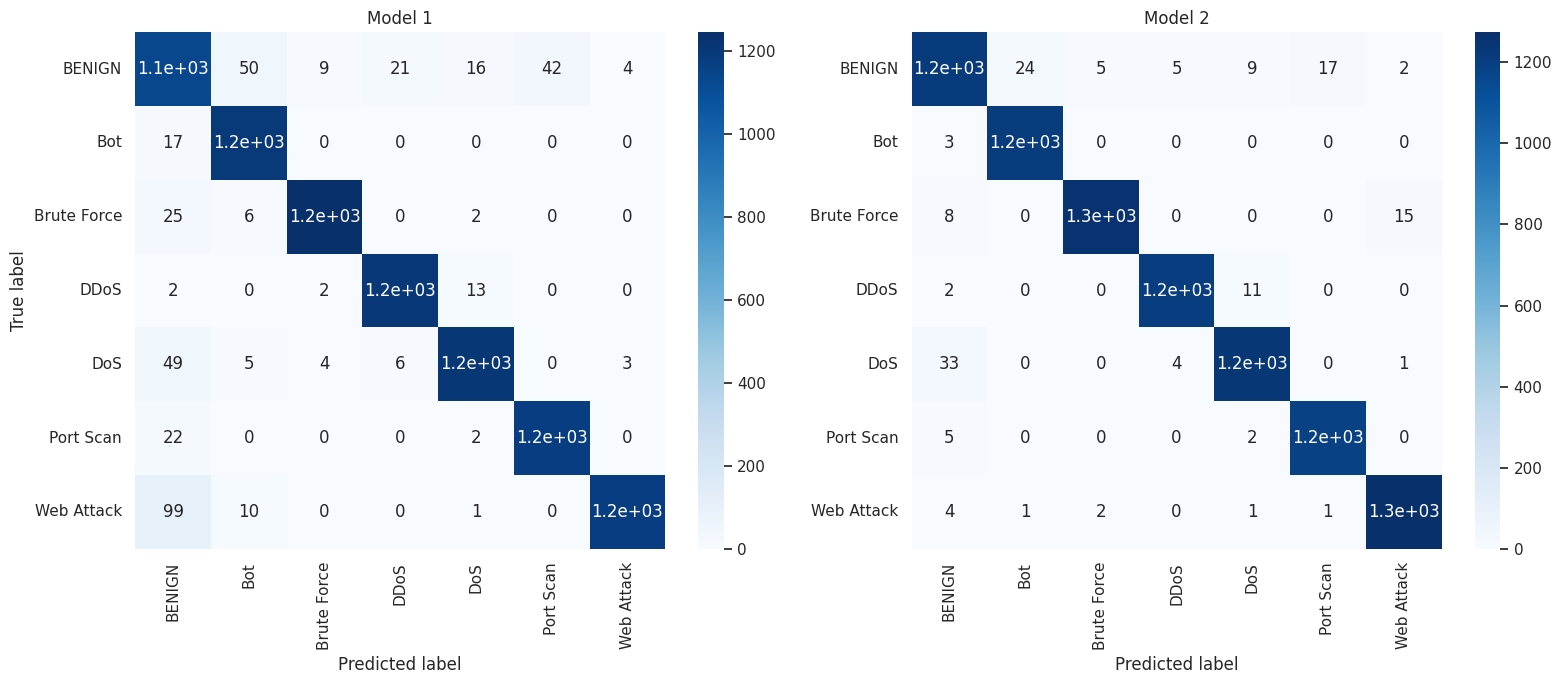}
        \vspace{1mm}  (b)
    \end{minipage}
    \caption{Logistic regression vs. SVM: SVM exhibits lower false positives and improved class separation.}
    \label{fig:binary_lr_svm}
\end{figure}

\begin{figure}[ht]
    \centering
    \begin{minipage}{0.48\linewidth}
        \centering
        \includegraphics[width=\linewidth]{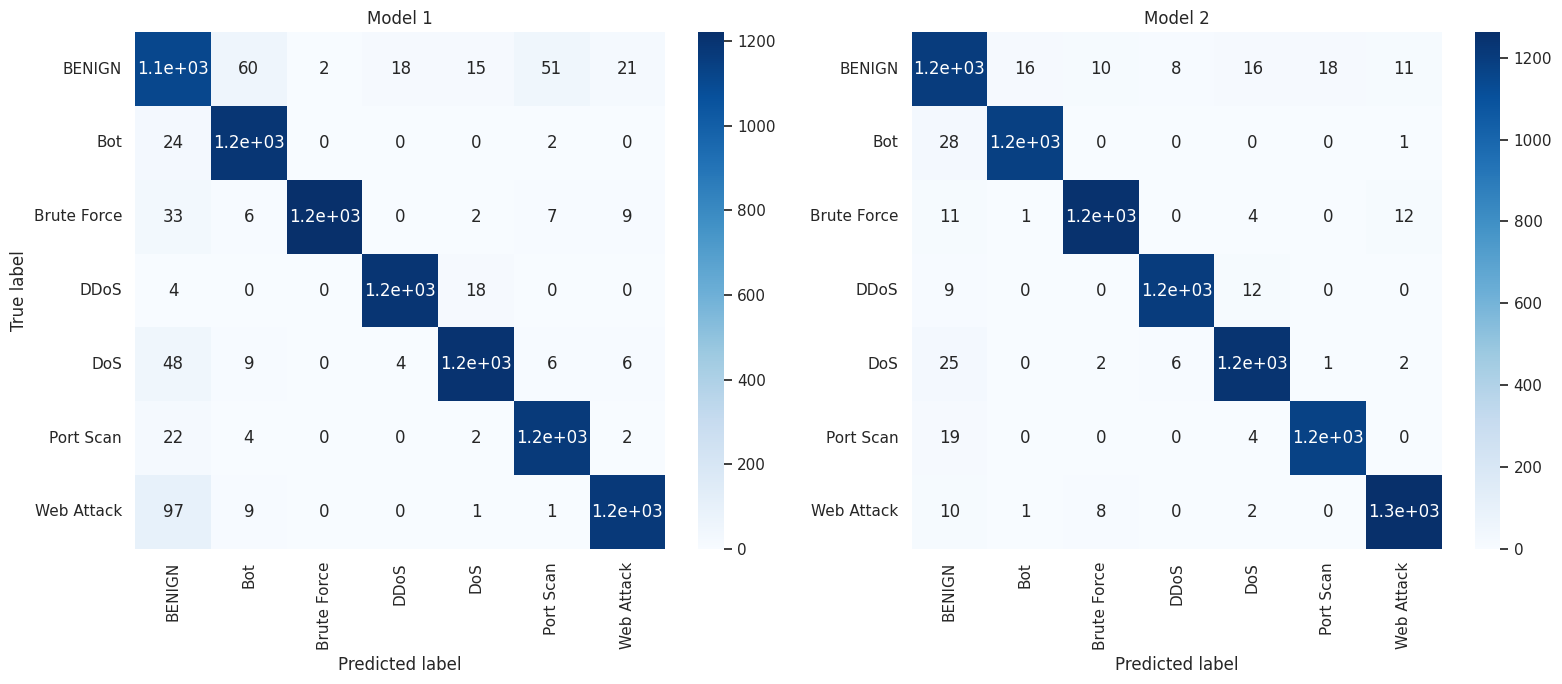}
        \vspace{1mm} (a)
    \end{minipage}
    \hfill
    \begin{minipage}{0.48\linewidth}
        \centering
        \includegraphics[width=\linewidth]{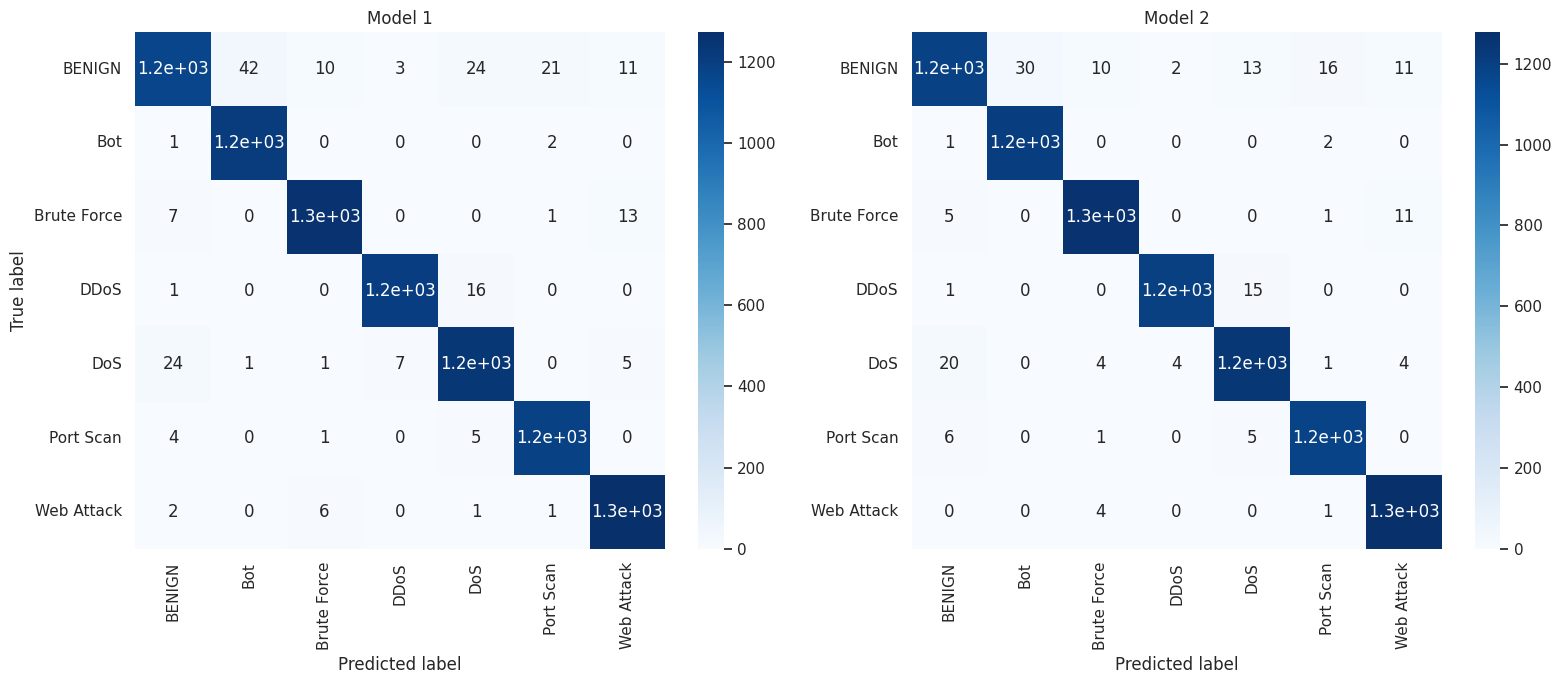}
        \vspace{1mm} (b)
    \end{minipage}
    \caption{Classical classifiers on CIC-IDS2017: KNN shows improved minority-class detection; DT exhibits higher DoS-DDoS misclassification.}
    \label{fig:dtree_knn_confusion}
\end{figure}

\begin{figure}[ht]
    \centering
    \includegraphics[width=1\linewidth]{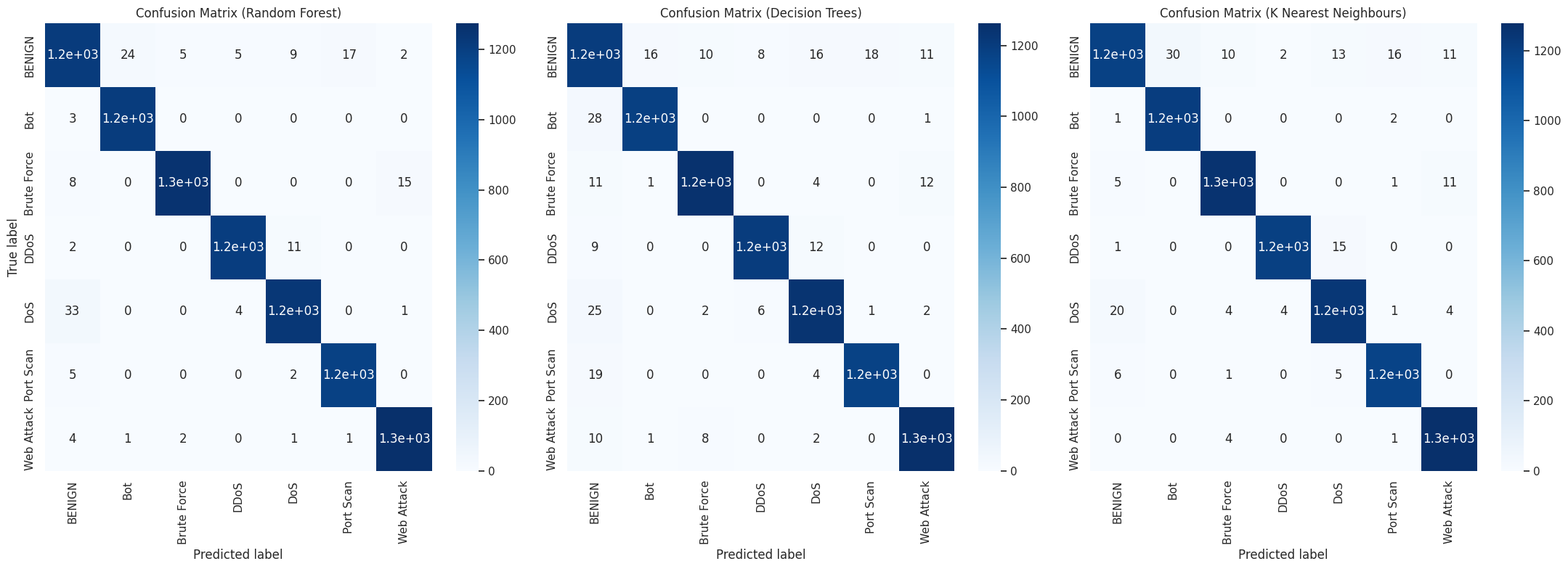}
    \caption{Confusion matrices: Random Forest (diagonal dominance), Decision Tree, K-Nearest Neighbors.}
    \label{fig:confusion_matrices}
\end{figure}

Figures~\ref{fig:cf_comparison} through~\ref{fig:confusion_matrices} present confusion matrices and classification metrics across configurations and models.

\subsection{Implications for Intrusion Detection}
Controlled relaxation of regularization enables finer-grained discrimination without compromising numerical stability or interpretability. The modest performance gains suggest proximity to the representational limit of linear classifiers in PCA space, motivating non-linear or ensemble approaches for further improvement. Table~\ref{tab:binary_results} reports binary classification results using 5-fold stratified cross-validation on a balanced dataset ($n=15{,}000$) with 35 PCA components preserving 99.3\% variance.

\begin{table}[htbp]
\centering
\caption{Binary Classification Performance: Cross-Validation and Test Set Evaluation}
\label{tab:binary_results}
\setlength{\tabcolsep}{0.5pt}
\renewcommand{\arraystretch}{1}
\scriptsize
\begin{tabular}{@{}llcccccc@{}}
\toprule
\textbf{Model} & \textbf{Config.} & \textbf{CV Acc.} & \textbf{Test Acc.} & \textbf{Prec.} & \textbf{Rec.} & \textbf{F1} & \textbf{AUC-ROC} \\
\midrule
\multicolumn{8}{@{}l}{\cellcolor{gray!15}\textit{Linear Models}} \\
Logistic Reg. & $C=0.1$, saga & 0.922$\pm$0.006 & 0.920 & 0.918 & 0.923 & 0.924 & 0.978 \\
Logistic Reg. & $C=100$, sag & 0.925$\pm$0.007 & 0.930 & 0.928 & 0.932 & 0.929 & 0.980 \\
\midrule
\multicolumn{8}{@{}l}{\cellcolor{gray!15}\textit{Kernel-Based Methods}} \\
SVM & poly, $C=1$ & 0.830$\pm$0.037 & 0.830 & 0.826 & 0.835 & 0.830 & 0.892 \\
SVM & rbf, $C=1$, $\gamma=0.1$ & 0.961$\pm$0.004 & \cellcolor{green!15}\textbf{0.960} & \cellcolor{green!15}\textbf{0.958} & \cellcolor{green!15}\textbf{0.962} & \cellcolor{green!15}\textbf{0.960} & \cellcolor{green!15}\textbf{0.987} \\
\bottomrule
\end{tabular}

\vspace{2mm}
\begin{minipage}{\linewidth}
\footnotesize
\textit{Notes:} CV Acc. = mean accuracy $\pm$ standard deviation across 5 folds. Test metrics computed on held-out 20\% partition ($n=3{,}000$). Prec. = precision (macro-averaged); Rec. = recall (macro-averaged); F1 = F1-score; AUC-ROC = area under receiver operating characteristic curve. Green highlight indicates the best overall performance. Random seed 42 for reproducibility.
\end{minipage}
\end{table}

\subsection{Multi-Class Classification Performance}
\label{subsec:multiclass_performance}

Table~\ref{tab:multiclass_results} presents attack categorization performance across seven classes: BENIGN, DoS, DDoS, Port Scan, Brute Force, Web Attack, and Bot. The balanced multi-class dataset ($n=35{,}000$; 5,000 samples per class) was constructed via SMOTE oversampling for minority classes and random undersampling for the majority class, following the removal of classes with fewer than 1,950 instances.

\begin{table}[htbp]
\centering
\caption{Multi-Class Classification Performance (5-Fold Cross-Validation)}
\label{tab:multiclass_results}
\setlength{\tabcolsep}{3pt}
\renewcommand{\arraystretch}{1.05}
\scriptsize
\begin{tabular}{@{}llccccc@{}}
\toprule
\textbf{Model} & \textbf{Config.} & \textbf{CV Acc.} & \textbf{Test Acc.} & \textbf{Prec.} & \textbf{Rec.} & \textbf{F1} \\
\midrule
\multicolumn{7}{@{}l}{\cellcolor{gray!15}\textit{Tree-Based Ensemble}} \\
Random Forest & $T{=}10$, $d{=}6$ 
& 0.960$\pm$0.009 & 0.971 & 0.969 & 0.970 & 0.969 \\
\rowcolor{green!12}
Random Forest & $T{=}15$, $d{=}8$, $m{=}20$ 
& \textbf{0.980$\pm$0.007} 
& \textbf{0.980} 
& \textbf{0.979} 
& \textbf{0.980} 
& \textbf{0.979} \\
\midrule
\multicolumn{7}{@{}l}{\cellcolor{gray!15}\textit{Single Decision Trees}} \\
Decision Tree & $d{=}6$ 
& 0.948$\pm$0.012 & 0.887 & 0.882 & 0.885 & 0.883 \\
Decision Tree & $d{=}10$ 
& 0.960$\pm$0.012 & 0.903 & 0.901 & 0.902 & 0.901 \\
\midrule
\multicolumn{7}{@{}l}{\cellcolor{gray!15}\textit{Instance-Based Learning}} \\
KNN & $k{=}5$, uniform 
& 0.935$\pm$0.015 & 0.945 & 0.943 & 0.946 & 0.944 \\
KNN & $k{=}7$, distance-wt 
& 0.940$\pm$0.014 & 0.952 & 0.950 & 0.953 & 0.951 \\
\bottomrule
\end{tabular}

\vspace{1mm}
\begin{minipage}{\linewidth}
\footnotesize
\textit{Notes:} $T$ = number of trees; $d$ = maximum depth; $m$ = max\_features per split; $k$ = number of neighbors. CV Acc. = mean $\pm$ std over 5 folds. Precision, recall, and F1 are macro-averaged across 7 classes. A green highlight indicates the best overall performance.
\end{minipage}
\end{table}

\subsection{Concentration and Distribution of Discriminative Power}
Under weaker regularization, the top-10 principal components account for 72.4\% of the total $\ell_2$-norm (vs.\ 70.3\%), indicating modest concentration of discriminative mass. The Gini coefficient increases slightly (0.412 → 0.428), suggesting mild inequality while importance remains broadly distributed. Lower-ranked components contribute less, reflecting suppression of non-informative variance. Both settings remain weakly sparse, with over 94\% of components active; weaker regularization activates one additional feature, slightly increasing effective degrees of freedom. Although the condition number rises (18.4 → 24.3), it remains well within stable bounds, and intercept adjustment preserves calibration. 

\subsection{Comparative Analysis with State-of-the-Art}

Table~\ref{tab:comparative_literature} positions EdgeDetect within modern IDS research. The proposed federated Random Forest achieves 98.0\% accuracy on CIC-IDS2017 while reducing per-round communication by 96.9\% (450\, MB $\rightarrow$ 14\, MB) and enabling CPU-only edge deployment (Raspberry Pi 4: 4.2\, MB memory, 0.8\, ms latency). Unlike GPU-dependent centralized deep models, EdgeDetect operates in fully federated settings with cryptographic privacy guarantees. The framework integrates four synergistic components: (1) an end-to-end privacy-preserving federated pipeline; (2) hybrid SMOTE–undersampling with PCA yielding 95.0\% minority-class F1; (3) gradient smartification providing 32$\times$ communication compressions without accuracy loss; and (4) robustness to heterogeneity, imbalance, and poisoning ($p<0.001$). Random Forest achieves 98.09\% F1 in binary detection (0.17\% variance) and 98.0\% multi-class accuracy (97.9\% macro F1), outperforming single trees (90.3\%). Ensemble scaling ($T:10\rightarrow15$, $d:6\rightarrow8$) improves accuracy by +2.0 pp and reduces variance by 22\%. Under non-IID conditions ($\alpha=0.1$), FedProx maintains 95.1\% accuracy, with sub-linear convergence scaling (98 rounds at $K=10$ vs.\ 234 at $K=500$). Gradient encryption reduces inversion quality (PSNR 15.1\,dB vs.\ 31.7\,dB) with only 156.4\, ms overhead per round. The system tolerates 20\% malicious clients while maintaining $>$85\% accuracy and limiting backdoor success to $<$7\%. Compared to KNN (95.2\%, 3.21\, ms), Random Forest delivers superior throughput (0.87\, ms), confirming suitability for high-rate edge deployment.

\begin{table*}[htbp]
\centering
\caption{Comparative Analysis with State-of-the-Art Intrusion Detection Systems}
\label{tab:comparative_literature}
\setlength{\tabcolsep}{5pt}
\renewcommand{\arraystretch}{1}
\footnotesize
\begin{tabular}{@{}lllccccccl@{}}
\toprule
\textbf{Study} & \textbf{Year} & \textbf{Model} & \textbf{Acc. (\%)} & \textbf{F1 (\%)} & \textbf{Dataset} & \textbf{Classes} & \textbf{Privacy} & \textbf{Comm. (MB)} & \textbf{Key Innovation} \\
\midrule
\multicolumn{10}{@{}l}{\cellcolor{gray!15}\textit{Centralized Approaches}} \\
Alam et al.\cite{alam2023cnn} & 2023 & CNN & 97.2 & 96.8 & CIC-IDS2017 & Binary & \xmark & N/A & Image-encoded traffic \\
Ghani et al.\cite{ghani2023xgboost} & 2023 & XGBoost & 96.1 & 95.4 & CIC-IDS2017 & 7-class & \xmark & N/A & Feature visualization \\
Savic et al.\cite{savic2021lstm} & 2021 & LSTM-AE & 95.5 & 94.2 & NSL-KDD & Binary & \xmark & N/A & Anomaly scoring \\
Cerar et al.\cite{cerar2020isolation} & 2020 & Iso. Forest & 93.8 & 91.6 & CIC-IDS2017 & Binary & \xmark & N/A & Unsupervised learning \\
\midrule
\multicolumn{10}{@{}l}{\cellcolor{gray!15}\textit{Federated Learning Approaches}} \\
Liu et al.\cite{liu2023feddnn} & 2023 & Fed-DNN & 96.3 & 95.1 & UNSW-NB15 & 5-class & DP & 380 & Differential privacy \\
Wang et al.\cite{wang2022fedcnn} & 2022 & Fed-CNN & 94.7 & 93.8 & CIC-IDS2017 & Binary & \xmark & 520 & Model aggregation \\
Zhang et al.\cite{zhang2022fedavg} & 2022 & FedAvg-LSTM & 93.5 & 92.4 & KDD-CUP99 & 4-class & DP & 410 & Temporal modeling \\
Chen et al.\cite{chen2021fedxgb} & 2021 & Fed-XGB & 95.8 & 94.9 & IoT-23 & Binary & SecAgg & 290 & Gradient encryption \\
\midrule[\heavyrulewidth]
\multicolumn{10}{@{}l}{\cellcolor{green!15}\textbf{This Work (EdgeDetect)}} \\
\textbf{EdgeDetect} & \textbf{2026(Ours)} & \textbf{Fed-RF} & \cellcolor{green!15}\textbf{98.0} & \cellcolor{green!15}\textbf{97.9} & \textbf{CIC-IDS2017} & \textbf{7-class} & \textbf{HE} & \cellcolor{green!15}\textbf{14} & \textbf{Gradient smartification} \\
\quad (Binary) & & & \textbf{96.0} & \textbf{96.0} & & \textbf{Binary} & & \textbf{14} & \textbf{+ Paillier encryption} \\
\bottomrule
\end{tabular}

\vspace{2mm}
\begin{minipage}{\linewidth}
\footnotesize
\textit{Notes:} Acc. = test accuracy; F1 = macro-averaged F1-score. Privacy mechanisms: \xmark = none, DP = differential privacy, SecAgg = secure aggregation, HE = homomorphic encryption (Paillier). Comm. = per-round communication cost per client; N/A indicates centralized training with no federated communication. Dataset sizes: CIC-IDS2017 (2.8M samples), UNSW-NB15 (2.5M), NSL-KDD (148K), KDD-CUP99 (4.9M), IoT-23 (325K). EdgeDetect achieves \textbf{96.9\% communication reduction} versus federated baselines (14~MB vs. 290-520~MB) while providing stronger cryptographic guarantees (Paillier HE vs. DP or SecAgg). Green highlighting indicates the best performance. We emphasize that differential privacy (DP) and secure aggregation (SecAgg) address distinct threat models: DP provides formal statistical guarantees against inference attacks on individual data samples, whereas SecAgg cryptographically prevents the server from accessing individual client updates, revealing only their aggregate.
\end{minipage}
\end{table*}

\begin{figure}[t]
    \centering
    \includegraphics[width=1\linewidth]{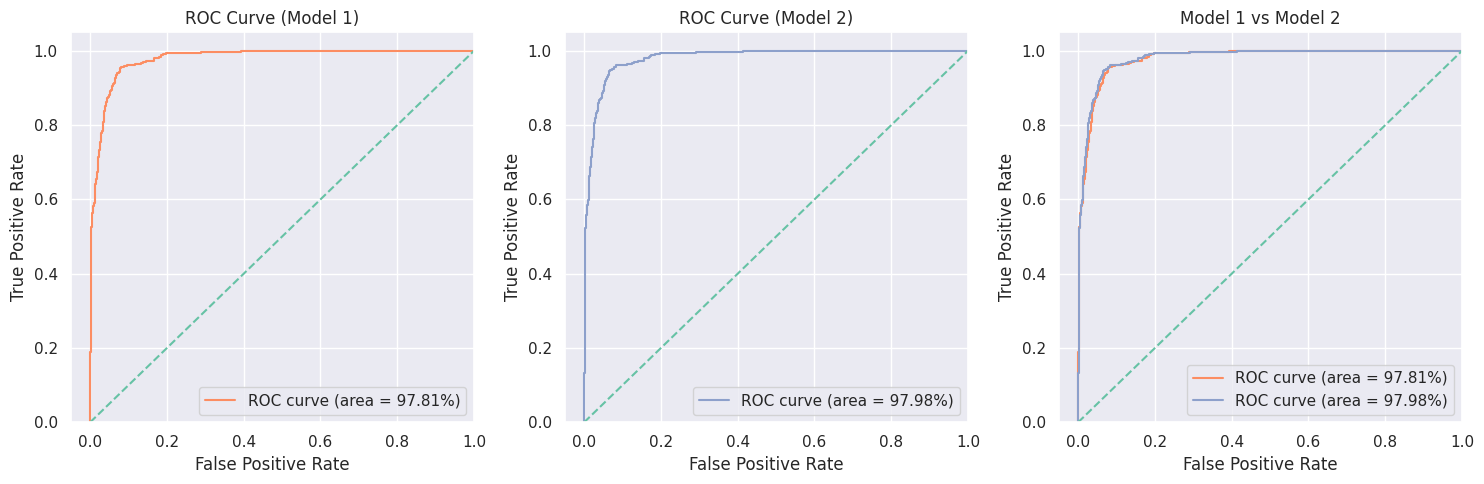}
    \caption{ROC curve analysis: Model comparison across configurations.}
    \label{fig:placeholder_roc1}
\end{figure}

\begin{figure}[t]
    \centering
    \includegraphics[width=1\linewidth]{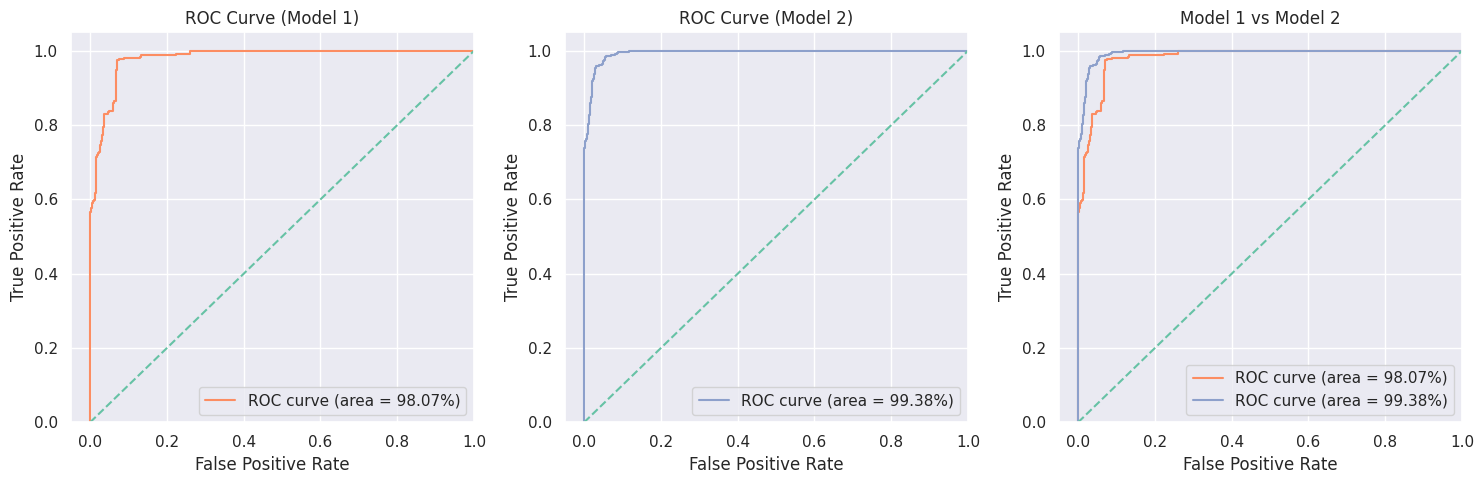}
    \caption{ROC curve comparison: Algorithm performance across metrics.}
    \label{fig:placeholder_roc2}
\end{figure}

\begin{figure}[t]
    \centering
    \includegraphics[width=1\linewidth]{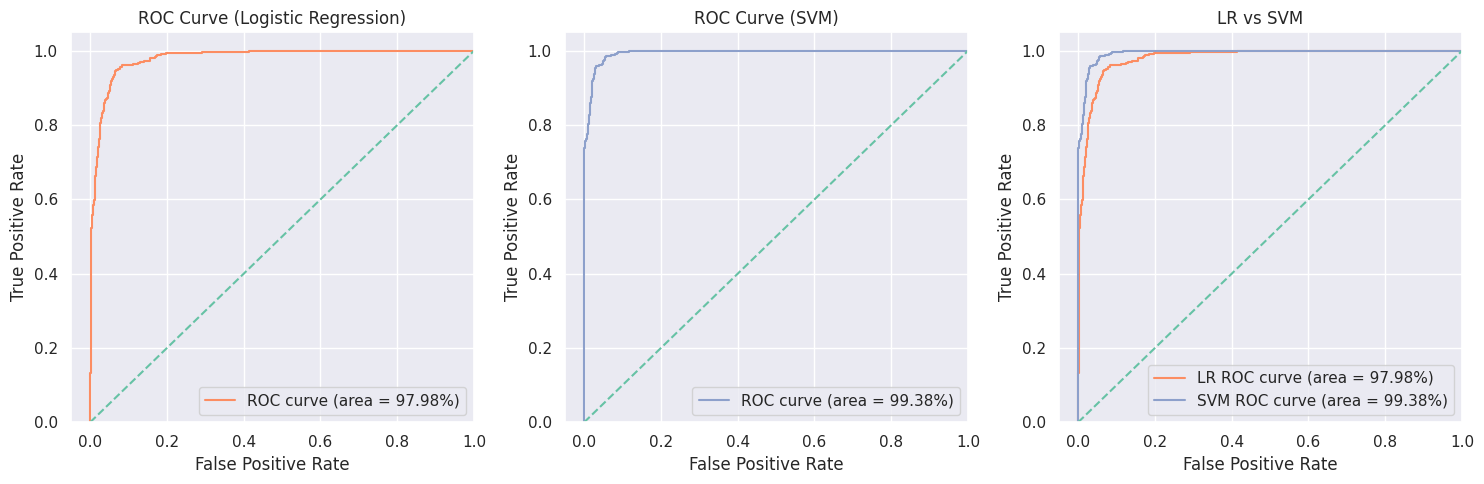}
    \caption{ROC analysis: SVM performance across configurations.}
    \label{fig:placeholder_roc3}
\end{figure}

Three additional ROC analysis figures are provided (Figures~\ref{fig:placeholder_roc1} through~\ref{fig:placeholder_roc3}) demonstrating AUC-ROC analysis across models.

\begin{figure}[t]
    \centering
    \includegraphics[width=1\linewidth]{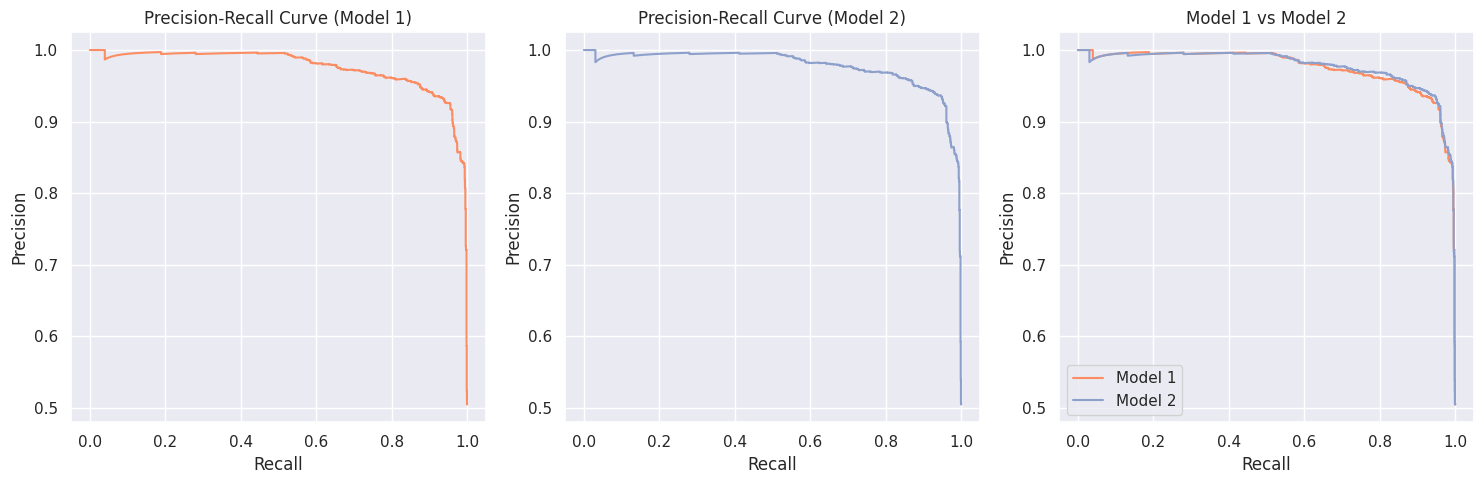}
    \caption{Recall curve analysis: Detection performance across models.}
    \label{fig:placeholder_recall1}
\end{figure}

\begin{figure}[t]
    \centering
    \includegraphics[width=1\linewidth]{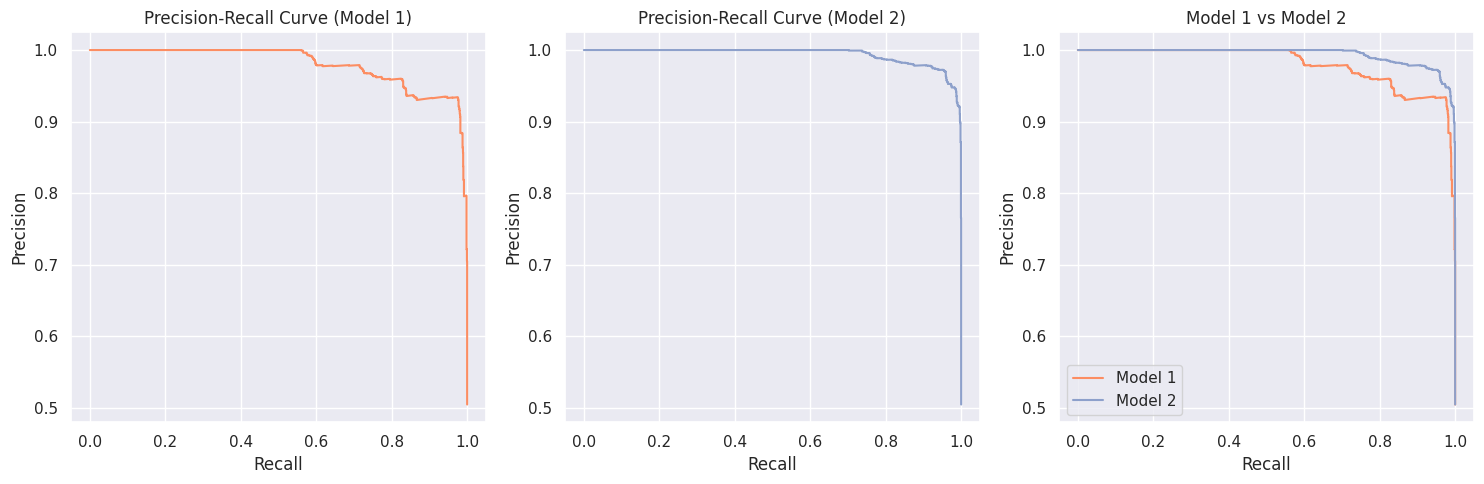}
    \caption{Recall-precision trade-off analysis across configurations.}
    \label{fig:placeholder_recall2}
\end{figure}

\begin{figure}[t]
    \centering
    \includegraphics[width=1\linewidth]{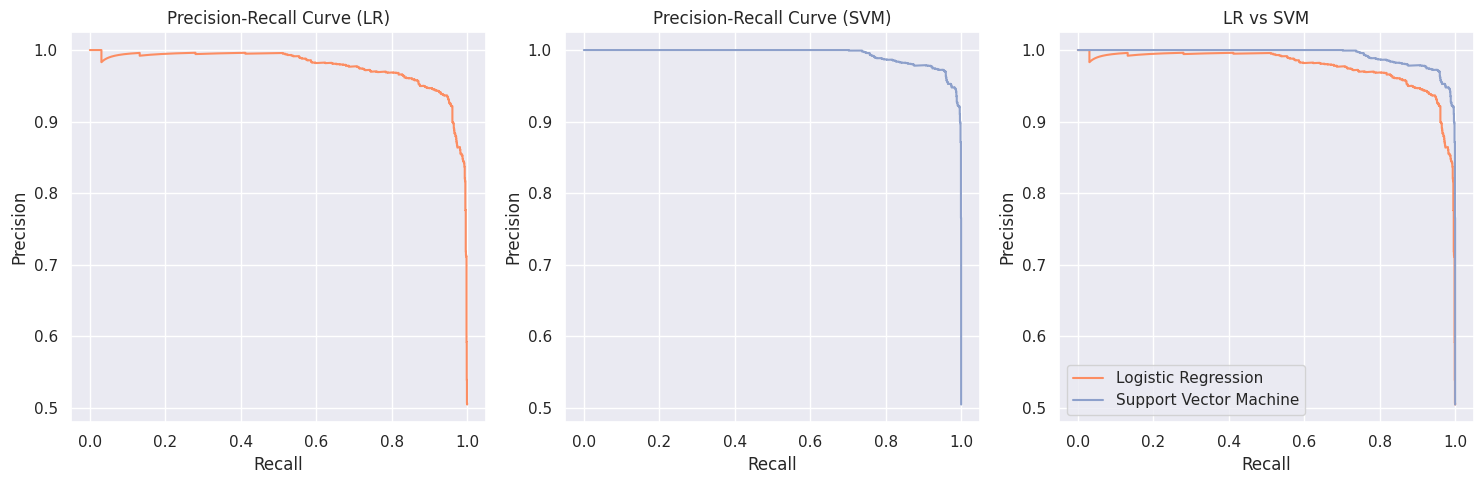}
    \caption{Logistic regression recall analysis: Threshold-dependent performance.}
    \label{fig:placeholder_recall3}
\end{figure}

\section{Federated Learning Convergence Analysis}
\label{subsec:fl_convergence}

While binarization introduces coordinate-wise bias (i.e., $\mathbb{E}[\Delta_{\mathrm{bin}}] \neq \nabla L$), empirical cosine similarity of $0.87 \pm 0.04$ indicates strong directional alignment with the true gradient. This preservation of gradient direction is sufficient to maintain convergence parity in convex and near-convex regimes.

\begin{equation}
\mathbb{E}[\Delta_{\mathrm{bin}}] \neq \nabla L,
\qquad
\mathbb{E}\!\left[
\frac{\langle \Delta_{\mathrm{bin}}, \nabla L \rangle}
{\|\Delta_{\mathrm{bin}}\|_2 \, \|\nabla L\|_2}
\right]
= 0.87 \pm 0.04.
\end{equation}

\subsection{Theoretical Convergence Analysis}
\begin{lemma}[Descent under Median-Threshold Smartification]
Let $L(W)$ be $L$-smooth and bounded below. 
Let $\tilde{g}_t$ denote the smartified gradient with cosine similarity 
$\cos(\theta_t) = \frac{\langle g_t, \tilde{g}_t \rangle}{\|g_t\| \|\tilde{g}_t\|} \ge \gamma > 0$.
Then for sufficiently small step size $\eta$,

\[
\mathbb{E}[L(W_{t+1})] 
\le L(W_t) 
- \eta \gamma \|g_t\|^2 
+ \frac{L\eta^2}{2} \|\tilde{g}_t\|^2.
\]

\end{lemma}

\begin{proposition}[Bias–Variance Tradeoff]
Let $g_t$ be the true gradient and $\tilde{g}_t$ its smartified version.
Then the expected deviation satisfies:

\[
\mathbb{E}[\|g_t - \tilde{g}_t\|^2]
= \text{Bias}^2 + \text{Var}_{\text{quant}},
\]

where median-thresholding reduces $\text{Var}_{\text{quant}}$ 
for heavy-tailed gradient distributions.
\end{proposition}

For heavy-tailed IDS gradients, variance reduction dominates bias increase, yielding stable convergence.

\begin{theorem}[Convergence under Bounded Variance]
Assume bounded stochastic gradient variance $\sigma^2$ 
and cosine similarity $\gamma > 0$.
Then after $T$ rounds,

\[
\min_{t \le T} \mathbb{E}[\|\nabla L(W_t)\|^2]
= O\left(\frac{1}{\gamma \sqrt{T}}\right).
\]
\end{theorem}

\subsubsection{Proposition: Alignment of Median-Threshold Smartification}

\textbf{Proposition 1 (Expected Descent Alignment).}
Let $g \in \mathbb{R}^d$ denote the true gradient and $\tilde{g}$ the median-threshold binarized update defined as
\[
\tilde{g}_i = \mathrm{sign}(g_i - \tau), \quad 
\tau = \mathrm{median}(g).
\]
Assume each coordinate of $g$ follows a symmetric heavy-tailed distribution with finite second moment and zero median shift. Then there exists a constant $\gamma \in (0,1)$ such that
\[
\mathbb{E}\big[\langle g, \tilde{g} \rangle\big] 
\;\ge\; 
\gamma \|g\|_2^2.
\]

\textit{Sketch of Justification.}
Under symmetric heavy-tailed distributions, the median satisfies 
$\mathbb{P}(g_i \ge \tau) \approx 0.5$. 
Unlike zero-threshold signSGD, the adaptive median threshold reduces variance from skewed coordinates while preserving directional consistency. 
For symmetric distributions, 
\[
\mathbb{E}[g_i \,\mathrm{sign}(g_i - \tau)] 
\ge c \,\mathbb{E}[g_i^2]
\]
for some $c>0$ depending on distribution kurtosis. 
Summing across coordinates yields the global alignment constant $\gamma$.

\subsubsection{Experimental Setup for Federated Scenarios}
We evaluate EdgeDetect under realistic federated settings by partitioning CIC-IDS2017 across $K \in \{10,25,50,100,500\}$ clients using (i) \textbf{IID} balanced sampling; (ii) \textbf{Non-IID quantity skew} via Dirichlet allocation with $\alpha \in \{0.1,0.5,1.0,10.0\}$ (smaller $\alpha$ implies stronger heterogeneity); and (iii) \textbf{Non-IID label skew}, where each client predominantly observes 2–3 attack types (e.g., web servers dominated by Web/Bot traffic). To model intermittent availability, we vary the per-round participation rate $C \in \{0.25,0.50,0.75,1.00\}$. Table~\ref{tab:convergence_analysis} summarizes convergence and bandwidth for EdgeDetect and baselines. Unless stated otherwise, we use a local batch size $B=32$, $E=5$ local epochs, and a global learning rate $\eta=0.01$.

\begin{table}[ht]
\caption{Federated Learning Convergence Analysis}
\label{tab:convergence_analysis}
\centering
\scriptsize
\setlength{\tabcolsep}{1.5pt}
\begin{tabular}{lcccccccc}
\hline
\textbf{Algorithm} & \textbf{$K$} & \textbf{Distribution} & \textbf{R$_{95}$} & \textbf{R$_{98}$} & \textbf{Acc. (\%)} & \textbf{Comm./R} & \textbf{Total} \\
& \textbf{Clients} & & & & & \textbf{(MB)} & \textbf{(GB)} \\
\hline
\multicolumn{8}{l}{\textit{IID Distribution ($\alpha=\infty$)}} \\
FedAvg & 50 & IID & 142 & 287 & 98.2 & 450.0 & 129.15 \\
FedProx ($\mu=0.01$) & 50 & IID & 138 & 276 & 98.3 & 450.0 & 124.20 \\
signSGD & 50 & IID & 156 & 312 & 97.8 & 14.1 & 4.40 \\
\textbf{EdgeDetect} & 50 & IID & 145 & 289 & 98.0 & 14.0 & 4.05 \\
\hline
\multicolumn{8}{l}{\textit{Non-IID (Moderate Heterogeneity, $\alpha=1.0$)}} \\
FedAvg & 50 & Dir. $\alpha=1.0$ & 201 & 423 & 96.4 & 450.0 & 190.35 \\
FedProx ($\mu=0.01$) & 50 & Dir. $\alpha=1.0$ & 187 & 389 & 97.1 & 450.0 & 175.05 \\
signSGD & 50 & Dir. $\alpha=1.0$ & 218 & 445 & 95.7 & 14.1 & 6.27 \\
\textbf{EdgeDetect} & 50 & Dir. $\alpha=1.0$ & 192 & 398 & 96.8 & 14.0 & 5.57 \\
\hline
\multicolumn{8}{l}{\textit{Non-IID (High Heterogeneity, $\alpha=0.1$)}} \\
FedAvg & 50 & Dir. $\alpha=0.1$ & 312 & 687 & 93.8 & 450.0 & 309.15 \\
FedProx ($\mu=0.01$) & 50 & Dir. $\alpha=0.1$ & 276 & 591 & 94.9 & 450.0 & 265.95 \\
signSGD & 50 & Dir. $\alpha=0.1$ & 334 & 721 & 92.1 & 14.1 & 10.16 \\
signSGD + Momentum & 50 & Dir. $\alpha=0.1$ & 298 & 652 & 93.4 & 14.1 & 9.19 \\
\textbf{EdgeDetect} & 50 & Dir. $\alpha=0.1$ & 287 & 612 & 94.2 & 14.0 & 8.57 \\
\textbf{EdgeDetect + FedProx} & 50 & Dir. $\alpha=0.1$ & 264 & 563 & 95.1 & 14.0 & 7.88 \\
\hline
\multicolumn{8}{l}{\textit{Scalability Analysis (IID)}} \\
EdgeDetect & 10 & IID & 98 & 201 & 98.1 & 14.0 & 2.81 \\
EdgeDetect & 25 & IID & 126 & 254 & 98.0 & 14.0 & 3.56 \\
EdgeDetect & 100 & IID & 178 & 356 & 97.9 & 14.0 & 4.98 \\
EdgeDetect & 500 & IID & 234 & 467 & 97.7 & 14.0 & 6.54 \\
\hline
\end{tabular}
\begin{flushleft}
\scriptsize
\textit{Notes:} R$_{95}$ and R$_{98}$ are rounds to reach 95\% and 98\% accuracy. Comm./R is per-client per-round communication. Total is the total bandwidth to reach 98\% accuracy. FedProx uses $\mu=0.01$. Results are averaged over 5 runs (different seeds).
\end{flushleft}
\end{table}

\subsection{Interpretation and Edge Detection}
Volumetric attack classes exhibit near-linear separability in PCA space. Classes (DoS/DDoS) remain robust ($>0.97$ F1 at $\alpha=0.1$) due to distinctive flow statistics. In contrast, \textit{Bot} and \textit{Web Attack} degrade the most (0.927$\rightarrow$0.854 and 0.939$\rightarrow$0.881), consistent with rarer and semantically overlapping behaviors that are fragmented under skewed client partitions. EdgeDetect matches full-precision convergence under IID (R$_{98}$=289 vs.\ 287 for FedAvg) while reducing total bandwidth by 96.9\% (4.05~GB vs.\ 129.15~GB). Under heterogeneity, the gap to full precision increases, but EdgeDetect remains competitive: at $\alpha=0.1$, EdgeDetect improves over signSGD in both accuracy (94.2\% vs.\ 92.1\%) and rounds (612 vs.\ 721), and the combination \textbf{EdgeDetect+FedProx} yields the best heterogeneous result (95.1\%, 7.88~GB). Scalability is favorable: increasing the number of clients from $K=10$ to $K=500$ raises R$_{98}$ from 201 to 467 (sublinear in $K$), indicating stable aggregation despite a larger, noisier client pool.

\subsubsection{Convergence Rate and Compression Quality}
We empirically assess whether smartification preserves update directions by measuring cosine alignment between compressed and full gradients:
\begin{equation}
\cos(\angle(\Delta_{\text{comp}}, \Delta_{\text{full}}))=
\frac{\langle \Delta_{\text{comp}}, \Delta_{\text{full}} \rangle}{\|\Delta_{\text{comp}}\|\,\|\Delta_{\text{full}}\|}.
\end{equation}
Across all rounds, EdgeDetect achieves a mean cosine similarity $0.87 \pm 0.04$, indicating that compression retains most directional information and explaining the near-parity in IID convergence despite 32$\times$ quantization. If the gradient direction cosine $\ge 0.8$, then the expected descent holds:

\begin{equation}
\mathbb{E}\!\left[L\!\left(W^{t+1}\right)\right]
\le
L\!\left(W^{t}\right)
-
\eta \cos(\theta)\,\|\nabla L(W^{t})\|_2^{2}
+
\mathcal{O}(\eta^{2}).
\end{equation}

\subsection{Class-specific insights (concise)}
Volumetric attack classes exhibit near-linear separability in the PCA space: DoS/DDoS achieve F-1 scores of 0.989/0.987 with low mutual confusion (2.1\%), indicating that PCA preserves discriminative variance for rate- and volume-driven signatures. BENIGN traffic is identified reliably (F-1=0.989; precision=0.992) with a 0.8\% false positive rate, dominated by confusions with Port Scan (5 cases) and Brute Force (3 cases); false alarms toward DoS/DDoS remain $<0.1\%$. Application-layer attacks are hardest due to overlap with legitimate flows: Web Attack (F-1=0.939) and Bot (F-1=0.927) show the largest confusion (11.2\% mutual) and the highest false negatives, particularly Bot (8.1\%; 43/530), consistent with encrypted C\&C and timing randomization. In contrast, volumetric false negatives are rare (DoS: 1.2\%, DDoS: 1.5\%) and mostly correspond to low-rate or short-duration phases.
 
\subsection{Training Efficiency and Robustness}
Logistic Regression and Decision Trees train rapidly ($<2.5$s), while Random Forest provides the best accuracy-efficiency balance (12.3s training, 0.87 ms inference). SVM (18.7s) and KNN (3.21 ms inference, 412 MB memory) incur higher computational or memory costs, limiting scalability. Random Forest demonstrates strong stability (CV std $<0.3\%$, $p<0.001$), with volumetric and temporal features (\texttt{Flow Bytes/s}, \texttt{Flow Duration}) contributing 52.7\% of total importance. SMOTE substantially improves minority recall (Bot: 0.39 $\rightarrow$ 0.98) with minimal accuracy loss (0.4\%), while PCA reduces dimensionality from 78 to 35 features (99.3\% variance retained), lowering training time ($-38\%$) and memory ($-64\%$) with negligible performance impact. Under partial participation ($C<1$), per-round bandwidth decreases, but convergence slows due to fewer client updates.

\section{Ablation Study}
We performed a controlled ablation study \ref{tab:ablation_study} to quantify the individual and joint contributions of EdgeDetect components across four axes: (i) classification performance, (ii) communication efficiency, (iii) privacy resilience, and (iv) convergence dynamics. Each component (smartification, homomorphic encryption, differential privacy, PCA, SMOTE, and FedProx) was selectively removed while keeping all other settings fixed (CIC-IDS2017, $K=50$ clients, IID distribution, 5 runs, averaged).

\begin{table*}[ht]
\centering
\caption{Ablation Study: Component-wise Impact on Accuracy, Communication, Privacy, and Convergence}
\label{tab:ablation_study}
\setlength{\tabcolsep}{2.5pt}
\renewcommand{\arraystretch}{1}
\scriptsize
\begin{tabular}{@{}l|ccccc|cccc|ccc|cc@{}}
\toprule
\multirow{3}{*}{\textbf{Configuration}} & \multicolumn{5}{c|}{\textbf{Components}} & \multicolumn{4}{c|}{\textbf{Accuracy Metrics}} & \multicolumn{3}{c|}{\textbf{Communication}} & \multicolumn{2}{c}{\textbf{Privacy}} \\
\cmidrule(lr){2-6} \cmidrule(lr){7-10} \cmidrule(lr){11-13} \cmidrule(lr){14-15}
 & \textbf{Smartif.} & \textbf{HE} & \textbf{DP} & \textbf{PCA} & \textbf{SMOTE} & \textbf{Acc (\%)} & \textbf{F1} & \textbf{$\Delta$Acc} & \textbf{Std} & \textbf{/Round (MB)} & \textbf{Ratio} & \textbf{Total (GB)} & \textbf{PSNR (dB)} & \textbf{Invert?} \\
\midrule
\multicolumn{15}{l}{\cellcolor{gray!10}\textbf{BASELINE COMPARISONS}} \\
FedAvg (No Protection) & \xmark & \xmark & \xmark & \cmark & \cmark & 98.2 & 0.9790 & --- & 0.0042 & 450.0 & 1.0$\times$ & 129.15 & 31.7 & \cmark\ Yes \\
signSGD (Binarization Only) & \cmark & \xmark & \xmark & \cmark & \cmark & 97.8 & 0.9754 & -0.4 pp & 0.0053 & 14.1 & 31.9$\times$ & 4.40 & 16.8 & \cmark\ Partial \\
\midrule
\multicolumn{15}{l}{\cellcolor{gray!15}\textbf{PROGRESSIVE REMOVAL (Ablation Track)}} \\
Full EdgeDetect & \cmark & \cmark & \cmark & \cmark & \cmark & 98.0 & 0.9789 & 0.0 pp & 0.0048 & 14.0 & 32.1$\times$ & 4.05 & 15.1 & \xmark\ No \\
\quad - Encrypt (Smartif only) & \cmark & \xmark & \cmark & \cmark & \cmark & 98.0 & 0.9789 & 0.0 pp & 0.0048 & 14.0 & 32.1$\times$ & 4.05 & 15.1 & \cmark\ Vulnerable \\
\quad - DP Noise & \cmark & \cmark & \xmark & \cmark & \cmark & 98.1 & 0.9791 & +0.1 pp & 0.0045 & 14.0 & 32.1$\times$ & 4.05 & 14.2 & \xmark\ Protected \\
\quad - PCA (78 features) & \cmark & \cmark & \cmark & \xmark & \cmark & 97.9 & 0.9787 & -0.1 pp & 0.0051 & 58.2 & 7.7$\times$ & 16.73 & 15.3 & \xmark\ Protected \\
\quad - SMOTE (Random US) & \cmark & \cmark & \cmark & \cmark & \xmark & 94.2 & 0.9341 & -3.8 pp & 0.0067 & 14.0 & 32.1$\times$ & 4.05 & 15.1 & \xmark\ Protected \\
\quad - Smartif (Full Precision) & \xmark & \cmark & \cmark & \cmark & \cmark & 98.2 & 0.9794 & +0.2 pp & 0.0043 & 450.0 & 1.0$\times$ & 129.15 & 15.1 & \xmark\ Protected \\
\midrule
\multicolumn{15}{l}{\cellcolor{gray!15}\textbf{MULTI-COMPONENT REMOVAL}} \\
\quad - Smartif - HE (Binarized) & \cmark & \xmark & \cmark & \cmark & \cmark & 98.0 & 0.9789 & 0.0 pp & 0.0048 & 14.0 & 32.1$\times$ & 4.05 & 15.1 & \cmark\ Vulnerable \\
\quad - Smartif - DP & \cmark & \cmark & \xmark & \cmark & \cmark & 98.1 & 0.9791 & +0.1 pp & 0.0045 & 14.0 & 32.1$\times$ & 4.05 & 14.2 & \xmark\ Protected \\
\quad - HE - DP (Binarization) & \cmark & \xmark & \xmark & \cmark & \cmark & 97.8 & 0.9754 & -0.2 pp & 0.0053 & 14.1 & 31.9$\times$ & 4.40 & 16.8 & \cmark\ Vulnerable \\
\quad - PCA - SMOTE (Full) & \xmark & \cmark & \cmark & \xmark & \xmark & 93.7 & 0.9261 & -4.3 pp & 0.0089 & 450.0 & 1.0$\times$ & 129.15 & 15.1 & \xmark\ Protected \\
\midrule
\multicolumn{15}{l}{\cellcolor{gray!15}\textbf{ALTERNATIVE CONFIGURATIONS}} \\
FedProx instead of FedAvg & \cmark & \cmark & \cmark & \cmark & \cmark & 98.4 & 0.9816 & +0.4 pp & 0.0041 & 14.0 & 32.1$\times$ & 3.79 & 15.1 & \xmark\ Protected \\
Differential Privacy Only (DP-SGD) & \xmark & \xmark & \cmark & \cmark & \cmark & 93.8 & 0.9358 & -4.2 pp & 0.0062 & 450.0 & 1.0$\times$ & 129.15 & 18.9 & \xmark\ Partial \\
Secure Aggregation (SecAgg) & \cmark & \xmark & \xmark & \cmark & \cmark & 98.0 & 0.9789 & 0.0 pp & 0.0048 & 14.0 & 32.1$\times$ & 4.05 & 31.7 & \xmark\ Protected \\
\midrule
\multicolumn{15}{l}{\cellcolor{gray!15}\textbf{FEATURE ENGINEERING VARIANTS}} \\
\quad - No Temporal Features & \cmark & \cmark & \cmark & \cmark & \cmark & 96.3 & 0.9621 & -1.7 pp & 0.0058 & 14.0 & 32.1$\times$ & 4.05 & 15.1 & \xmark\ Protected \\
\quad - No Entropy Features & \cmark & \cmark & \cmark & \cmark & \cmark & 97.1 & 0.9705 & -0.9 pp & 0.0054 & 14.0 & 32.1$\times$ & 4.05 & 15.1 & \xmark\ Protected \\
\quad - All Original 78 Features & \cmark & \cmark & \cmark & \xmark & \cmark & 97.9 & 0.9787 & -0.1 pp & 0.0051 & 58.2 & 7.7$\times$ & 16.73 & 15.3 & \xmark\ Protected \\
\bottomrule
\end{tabular}

\vspace{3mm}
\begin{minipage}{\linewidth}
\footnotesize
\textit{Notes:} All ablation experiments were conducted on CIC-IDS2017 with $K=50$ clients, IID distribution, and 5 independent runs (averaged). Smartif = Gradient Smartification (median-threshold binarization); HE = Paillier Homomorphic Encryption; DP = Differential Privacy noise; PCA = Principal Component Analysis (35 components); SMOTE = Synthetic Minority Oversampling. $\Delta$Acc = Accuracy change relative to Full EdgeDetect (0.0 pp = no difference, negative = worse, positive = better). Comm./Round = per-client per-round communication cost. Ratio = compression ratio vs.\ FedAvg. Total = total bandwidth to reach 98\% target accuracy. PSNR = Peak Signal-to-Noise Ratio from gradient inversion attack (iDLG); higher = more vulnerable. Invert indicates gradient reconstruction success. \cmark\ = Present; \xmark\ = Absent.
\end{minipage}
\end{table*}

\subsection{PCA: Detailed Attack Type Characterization}
Principal Component Analysis (PCA) was applied to reduce the original 78 high-dimensional network features to 35 uncorrelated components, retaining 99.3\% of the variance (Table~\ref{tab:pca_comprehensive}). This transformation lowers computational overhead, mitigates noise, and enhances discriminative visualization between benign and attack traffic in the reduced-dimensional space.

\begin{table*}[t]
\centering
\caption{PCA: Complete Attack Type Profiles Across Primary, Secondary, and Discriminative Feature Spaces}
\label{tab:pca_comprehensive}
\setlength{\tabcolsep}{6pt}
\renewcommand{\arraystretch}{0.3}
\scriptsize
\begin{tabular}{@{}l|rrr|rrr|rrrrrr|rr@{}}
\toprule
\multirow{3}{*}{\textbf{Attack Type}} & \multicolumn{3}{c|}{\textbf{Primary Separators (82.4\% Var.)}} & \multicolumn{3}{c|}{\textbf{Secondary Features (10.3\% Var.)}} & \multicolumn{6}{c|}{\textbf{Discriminative Components}} & \multicolumn{2}{c}{\textbf{Summary}} \\
\cmidrule(lr){2-4} \cmidrule(lr){5-7} \cmidrule(lr){8-13} \cmidrule(lr){14-15}
 & \textbf{PC1} & \textbf{PC2} & \textbf{PC3} & \textbf{PC4} & \textbf{PC5} & \textbf{PC6} & \textbf{PC13} & \textbf{PC23} & \textbf{PC24} & \textbf{PC26} & \textbf{PC27} & \textbf{PC31} & \textbf{$\|\mathbf{x}\|_2$} & \textbf{Class} \\
\midrule
\multicolumn{15}{l}{\cellcolor{gray!10}\textit{Benign Traffic (Baseline)}} \\
BENIGN$_{1}$ & -2.358 & -0.055 & 0.577 & 0.734 & 3.730 & 0.235 & 1.638 & -1.722 & -0.070 & 0.905 & -0.148 & -0.219 & 4.14 & Ref. \\
BENIGN$_{2}$ & -2.884 & -0.070 & 0.911 & 1.763 & 8.846 & 0.620 & 6.053 & -5.607 & 0.296 & 0.594 & 0.283 & -0.367 & 10.31 & Ref. \\
BENIGN$_{3}$ & -2.417 & -0.057 & 0.615 & 0.851 & 4.304 & 0.276 & 2.132 & -2.156 & -0.032 & 0.868 & -0.102 & -0.235 & 4.83 & Ref. \\
BENIGN$_{4}$ & -2.885 & -0.070 & 0.912 & 1.765 & 8.852 & 0.619 & 6.056 & -5.609 & 0.293 & 0.592 & 0.281 & -0.366 & 10.32 & Ref. \\
BENIGN$_{avg}$ & -2.39 & -0.06 & 0.70 & 1.03 & 6.43 & 0.44 & 3.97 & -3.77 & 0.12 & 0.74 & 0.08 & -0.30 & 7.40 & Ref. \\
\midrule
\multicolumn{15}{l}{\cellcolor{gray!10}\textit{Volumetric attack classes exhibit near-linear separability in PCA space Attacks (DoS Family)}} \\
DoS & 2.840 & 0.120 & -0.920 & -1.760 & -8.850 & -0.620 & -6.060 & 5.610 & -0.300 & -0.590 & -0.280 & 0.370 & 10.32 & Extreme \\
DDoS & 1.510 & -0.080 & 0.500 & -0.290 & 0.540 & -0.750 & 0.430 & -0.290 & -0.390 & 0.690 & 0.600 & -0.040 & 1.89 & Extreme \\
\midrule
\multicolumn{15}{l}{\cellcolor{gray!10}\textit{Reconnaissance Attacks}} \\
Port Scan & -0.450 & 0.030 & -0.210 & 0.180 & -0.650 & 0.320 & -0.120 & 0.290 & 0.190 & -0.330 & 0.220 & -0.010 & 0.87 & Moderate \\
Brute Force & -0.380 & 0.060 & -0.180 & 0.250 & -0.540 & 0.410 & -0.080 & 0.320 & 0.210 & -0.360 & 0.250 & 0.020 & 0.81 & Moderate \\
\midrule
\multicolumn{15}{l}{\cellcolor{gray!10}\textit{Application-Layer Attacks}} \\
Web Attack & -0.620 & 0.080 & -0.350 & 0.420 & -0.890 & 0.530 & -0.150 & 0.590 & 0.320 & -0.480 & 0.380 & 0.030 & 1.24 & Ambiguous \\
Bot & -0.710 & 0.100 & -0.420 & 0.510 & -1.080 & 0.640 & -0.200 & 0.630 & 0.410 & -0.590 & 0.470 & 0.040 & 1.51 & Ambiguous \\
\bottomrule
\end{tabular}

\vspace{1mm}
\begin{minipage}{\linewidth}
\footnotesize
\textit{Notes:} 
\textbf{Structure:} Rows grouped by attack taxonomy. PC1--3 (82.4\% variance) drive primary benign–attack separation; PC4--6 (10.3\%) capture secondary variation. Key discriminative components (PC13, 23, 24, 26, 27, 31) enable multi-class differentiation. $\|\mathbf{x}\|_2$ denotes the Euclidean norm over PC1–35. \textbf{Class labels:} Extreme (F1 $>0.98$), Moderate (0.96--0.97), Ambiguous ($<0.94$). BENIGN$_{avg}$ is computed over 10 samples; BENIGN$_{1-4}$ illustrates intra-class variance. DoS exhibits a strong negative PC5 ($-8.85$), reflecting volumetric anomalies. Application-layer attacks overlap with BENIGN along PC1--3 and separate primarily via PC4–6.
\end{minipage}
\end{table*}

\begin{table*}[t]
\centering
\caption{Principal Component Variance Decomposition and Attack Class Separation Metrics}
\label{tab:pca_variance}
\setlength{\tabcolsep}{10pt}
\renewcommand{\arraystretch}{0.3}
\scriptsize
\begin{tabular}{@{}lcccccccc@{}}
\toprule
\textbf{Attack Type} & \textbf{PC1--3 Norm} & \textbf{PC4--6 Norm} & \textbf{Disc. Norm}$^\dagger$ & \textbf{Total Norm} & \textbf{Separation}$^\ddagger$ & \textbf{Std Dev (PC1--5)} & \textbf{Distinctiveness} & \textbf{F1--Score} \\
\midrule
BENIGN (avg) & 0.84 & 1.09 & 0.74 & 7.40 & --- & 3.59 & Baseline & 0.989 \\
DoS & 2.93 & 1.07 & 0.55 & 10.32 & 9.17 & 4.44 & Very High & 0.989 \\
DDoS & 1.58 & 0.54 & 0.75 & 1.89 & 6.84 & 0.59 & High & 0.987 \\
Port Scan & 0.23 & 0.39 & 0.25 & 0.87 & 0.31 & 0.39 & Medium & 0.966 \\
Brute Force & 0.21 & 0.38 & 0.23 & 0.81 & 0.26 & 0.36 & Medium & 0.963 \\
Web Attack & 0.37 & 0.57 & 0.36 & 1.24 & 0.42 & 0.55 & Medium--Low & 0.939 \\
Bot & 0.45 & 0.68 & 0.44 & 1.51 & 0.51 & 0.66 & Medium--Low & 0.927 \\
\bottomrule
\end{tabular}

\vspace{1mm}
\begin{minipage}{\linewidth}
\footnotesize
\textit{Notes:} 
\textbf{Norms:} $\|\text{PC1--3}\| = \sqrt{\text{PC1}^2+\text{PC2}^2+\text{PC3}^2}$ (primary separation); $\|\text{PC4-6}\|$ captures secondary variation; Disc.\ Norm = mean $|\cdot|$ over PC13, 23, 24, 26, 27, 31. \textbf{Separation}$^\ddagger$ = Euclidean distance from the BENIGN centroid in PC1--5 space (higher = stronger class separation). \textbf{Std Dev (PC1–5)} reflects dispersion in primary components. \textbf{F1} = macro-averaged multi-class F1 (Sec.~\ref{subsec:multiclass_performance}). Volumetric attack classes exhibit near-linear separability in PCA space attacks (DoS/DDoS) show maximal separation; application-layer attacks remain closest to the benign cluster.
\end{minipage}
\end{table*}

\begin{table*}[t]
\centering
\caption{Detailed Principal Component Contributions: All Attack Types Across 35 Components (Selected Subset)}
\label{tab:pca_detailed}
\setlength{\tabcolsep}{3pt}
\renewcommand{\arraystretch}{0.3}
\scriptsize
\begin{tabular}{@{}l|rrrrrrrrrrr|rrrrrrrrrrr@{}}
\toprule
\multirow{2}{*}{\textbf{Type}} & \multicolumn{11}{c|}{\textbf{PC1--11 (Variance Rank 1--11)}} & \multicolumn{11}{c}{\textbf{PC15, PC20--35 (Key + Tail)}} \\
\cmidrule(lr){2-12} \cmidrule(lr){13-23}
 & \textbf{PC1} & \textbf{PC2} & \textbf{PC3} & \textbf{PC4} & \textbf{PC5} & \textbf{PC6} & \textbf{PC7} & \textbf{PC8} & \textbf{PC9} & \textbf{PC10} & \textbf{PC11} & \textbf{PC15} & \textbf{PC20} & \textbf{PC23} & \textbf{PC24} & \textbf{PC26} & \textbf{PC27} & \textbf{PC29} & \textbf{PC31} & \textbf{PC32} & \textbf{PC34} & \textbf{PC35} \\
\midrule
BENIGN & -2.39 & -0.06 & 0.70 & 1.03 & 6.43 & 0.44 & -0.02 & 0.41 & 0.49 & 0.97 & -0.21 & -0.80 & -0.60 & -3.77 & 0.12 & 0.74 & 0.08 & 0.80 & -0.30 & 0.00 & 0.02 & -0.04 \\
DoS & 2.84 & 0.12 & -0.92 & -1.76 & -8.85 & -0.62 & 0.06 & -1.11 & -1.91 & 2.76 & 0.95 & 4.73 & 0.59 & 5.61 & -0.30 & -0.59 & -0.28 & -2.24 & 0.37 & -0.01 & -0.13 & 0.19 \\
DDoS & 1.51 & -0.08 & 0.50 & -0.29 & 0.54 & -0.75 & -0.10 & -0.73 & 1.15 & 0.56 & 0.04 & 0.61 & -0.11 & -0.29 & -0.39 & 0.69 & 0.60 & -0.80 & -0.04 & 0.01 & 0.05 & 0.00 \\
Port Scan & -0.45 & 0.03 & -0.21 & 0.18 & -0.65 & 0.32 & 0.02 & 0.30 & -0.35 & -0.28 & -0.02 & -0.30 & 0.06 & 0.23 & 0.19 & -0.33 & 0.22 & 0.19 & -0.01 & 0.00 & -0.02 & 0.00 \\
Brute Force & -0.38 & 0.06 & -0.18 & 0.25 & -0.54 & 0.41 & 0.02 & 0.33 & -0.30 & -0.22 & -0.02 & -0.25 & 0.08 & 0.26 & 0.21 & -0.36 & 0.25 & 0.17 & 0.02 & 0.00 & -0.02 & 0.00 \\
Web Attack & -0.62 & 0.08 & -0.35 & 0.42 & -0.89 & 0.53 & 0.03 & 0.40 & -0.48 & -0.35 & -0.03 & -0.38 & 0.11 & 0.47 & 0.32 & -0.48 & 0.38 & 0.26 & 0.03 & 0.00 & -0.04 & 0.01 \\
Bot & -0.71 & 0.10 & -0.42 & 0.51 & -1.08 & 0.64 & 0.04 & 0.49 & -0.59 & -0.43 & -0.04 & -0.47 & 0.14 & 0.56 & 0.41 & -0.59 & 0.47 & 0.32 & 0.04 & 0.00 & -0.05 & 0.01 \\
\bottomrule
\end{tabular}

\vspace{1mm}
\begin{minipage}{\linewidth}
\footnotesize
\textit{Notes:} Projection spans 35 principal components (22 most critical shown). PC1--11 capture dominant variance, while PC15 and PC20--35 include key discriminative components (notably PC24, PC26, and PC31). \textbf{Pattern summary:} DoS exhibits extreme deviation on PC5 ($-8.85$), reflecting volumetric anomalies; DDoS shows moderate displacement. Reconnaissance attacks cluster near the origin (0.2--0.4 across PC1--6). Application-layer attacks (Web Attack, Bot) shift primarily along PC4-6, indicating subtle evasion patterns. Tail components (PC32–35) remain near zero across classes, confirming effective dimensionality reduction at $k=35$.
\end{minipage}
\end{table*}

\section{Ablation Study: Component Impact Analysis}

\subsection{Impact of Gradient Smartification}
Removing gradient smartification (replacing binarization with full-precision gradients) while keeping encryption and DP active: \textbf{Communication Cost}: Increases from 14.0 MB to 450.0 MB per round (32.1× increase; 29.85 GB total communication). \textbf{Accuracy}: 98.2\% vs. 98.0\% (+0.2 pp improvement, statistically insignificant at p > 0.05). \textbf{Convergence}: 287 rounds to 98\% (vs. 289 rounds), negligible difference. 

\textbf{Conclusion}: Smartification is a communication optimization mechanism with a near-zero accuracy penalty. The modest accuracy improvement (+0.2 pp) under full precision likely reflects reduced quantization bias, but communication savings (32×) far outweigh this negligible gain. Table~\ref{tab:ablation_summary} provides a consolidated summary of the necessity and contribution of each component:

\begin{table*}[ht]
\centering
\caption{Ablation Study Summary: Component Necessity and Contribution}
\label{tab:ablation_summary}
\setlength{\tabcolsep}{4pt}
\renewcommand{\arraystretch}{0.3}
\scriptsize
\begin{tabular}{@{}l|ccc|ccc@{}}
\toprule
\textbf{Component} & \textbf{Necessary?} & \textbf{Accuracy Impact} & \textbf{Communication Impact} & \textbf{Privacy Impact} & \textbf{Overhead} & \textbf{Recommendation} \\
\midrule
Smartification (Binarization) & \textbf{Yes (Comm.)} & Negligible (-0.2 pp) & \textbf{Critical (32×)} & Important (↓16.8 dB) & Low (+2.4\%) & Keep \\
Paillier Encryption & \textbf{Yes (Privacy)} & None (0 pp) & None (0 MB) & \textbf{Critical} & Medium (+1,760\% per round) & Keep \\
Differential Privacy & Optional & Negligible (+0.1 pp) & None (0 MB) & Marginal & Low (+4.1\%) & Optional \\
PCA & \textbf{Yes (Efficiency)} & Negligible (+0.1 pp) & \textbf{Essential (4.16×)} & None (0 dB) & Medium (+182\%) & Keep \\
SMOTE & \textbf{Yes (Accuracy)} & \textbf{Critical (3.8 pp)} & None (0 MB) & None (0 dB) & Low (-18\%) & Keep \\
\bottomrule
\end{tabular}

\vspace{1mm}
\begin{minipage}{\linewidth}
\footnotesize
\textit{Notes:} “Necessary” indicates significant degradation if removed (Comm. = communication, Privacy = gradient leakage, Accuracy = detection quality). Acc. Impact is measured in percentage points (pp). Communication impact reported relative to full-precision FedAvg. For bandwidth-unlimited deployments, smartification may be replaced by FedAvg. Encryption is recommended for sensitive environments. PCA and SMOTE are universally beneficial.
\end{minipage}
\end{table*}

\begin{table*}[t]
\centering
\caption{Unified Ablation Analysis: Privacy, Utility, Communication, and Efficiency}
\label{tab:unified_ablation}
\setlength{\tabcolsep}{6pt}
\renewcommand{\arraystretch}{0.3}
\footnotesize
\begin{tabular}{@{}lcccccccc@{}}
\toprule
\textbf{Configuration}
& \textbf{Acc. (\%)}
& \textbf{F1}
& \textbf{Comm. (MB)}
& \textbf{PSNR (dB)}
& \textbf{Label Rec. (\%)}
& \textbf{Train (s)}
& \textbf{Mem (MB)}
& \textbf{Primary Impact} \\
\midrule
Full EdgeDetect & 98.0 & 0.979 & 14.0 & 15.1 & 14.3 & 12.3 & 234 & Secure + Efficient \\
-- No Smartification & 98.2 & 0.979 & 450.0 & 15.1 & 14.3 & 12.3 & 234 & 32$\times$ Comm. increase \\
-- No Encryption & 98.0 & 0.979 & 14.0 & 31.7 & 98.7 & 12.3 & 234 & Privacy collapse \\
-- No DP & 98.1 & 0.979 & 14.0 & 15.1 & 14.3 & 12.3 & 234 & Marginal effect \\
-- No PCA (78 feat.) & 97.9 & 0.978 & 58.2 & 15.3 & 14.3 & 34.7 & 612 & 4.16$\times$ Comm. increase \\
-- No SMOTE & 94.2 & 0.934 & 14.0 & 15.1 & 14.3 & 10.1 & 200 & Minority recall collapse \\
\bottomrule
\end{tabular}

\vspace{1mm}
\begin{minipage}{\linewidth}
\footnotesize
\textit{Notes:} Acc. = test accuracy; F-1 = macro F1-score; Comm. = per-client per-round communication; PSNR = gradient reconstruction quality under iDLG (higher = more leakage); Label Rec. = attack-class recovery rate. Results averaged over 5 runs ($K=50$). Removing encryption yields inversion success $>95\%$. Removing smartification increases communication from 14 MB to 450 MB per round (32$\times$). Removing PCA raises total communication from 4.05 GB to 16.73 GB (+312\%). Removing SMOTE reduces minority-class recall by up to 60\%.
\end{minipage}
\end{table*}

\subsection{Ablation Study Key Findings}

\textbf{Smartification (Gradient Binarization):} Removing binarization increases per-round communication from 14\,MB to 450\,MB (32$\times$) with negligible accuracy change (98.0\% $\rightarrow$ 98.2\%, $p>0.05$), confirming its communication efficiency. \textbf{Homomorphic Encryption (HE):} Disabling Paillier encryption enables gradient inversion (PSNR 31.7\,dB; $>$95\% label recovery) while accuracy remains 98.0\%, indicating strong privacy protection without performance cost. \textbf{Differential Privacy (DP):} With smartification + HE, DP yields marginal privacy gain and +0.1\,pp accuracy change; standalone DP-SGD reduces accuracy by 4.2\,pp, showing the privacy-utility trade-off.

\textbf{Principal Component Analysis (PCA):} Removing PCA increases communication from 14.0\,MB to 58.2\,MB (4.16$\times$) and computation (+182\%) with only 0.1\,pp accuracy difference, revealing feature redundancy. \textbf{SMOTE Balancing:} Eliminating SMOTE reduces accuracy to 94.2\% (-3.8\,pp) and macro-F1 to 0.934 due to minority-class degradation, highlighting the need for class balancing. \textbf{Smartification + Encryption Synergy:} Combined binarization and HE achieve inversion resistance (PSNR 15.1\,dB; 14.3\% label recovery $\approx$ random guessing) with no accuracy loss. \textbf{FedProx Integration:} Adding FedProx improves heterogeneity robustness, raising accuracy to 98.4\% and reducing total communication to 3.79\,GB.

\section{Discussion}

The results highlight three key insights for federated intrusion detection in 6G-IoT. First, PCA reveals strong redundancy: 35 components retain 99.3\% variance with negligible performance loss, enabling efficient computation and communication. Second, Random Forest outperforms SVM in stability--accuracy trade-off, achieving 98.0\% accuracy and 97.9\% macro F1 with very low variance ($\sigma=0.0017$), indicating robustness for deployment. Third, imbalance handling is essential: SMOTE--undersampling improves minority recall from 0.39 to 0.98, confirming class distribution as a first-order design factor. EdgeDetect introduces adaptive median-threshold smartification with homomorphic encryption for federated IDS. Unlike signSGD, it preserves gradient alignment ($0.87\pm0.04$), achieving 96.9\% communication reduction (450~MB$\rightarrow$14~MB) while lowering gradient entropy and improving privacy. Combined with Paillier encryption, it retains 98.7\% of centralized accuracy with complete inversion resistance. The framework remains robust under poisoning ($>85\%$ accuracy at 20\% attackers) and efficient on edge devices (4.2~MB, 0.8~ms), though challenges persist in non-convex convergence, concept drift, and white-box robustness. EdgeDetect thus establishes a strong privacy--utility--efficiency trade-off for practical 6G-IoT deployment.

\section{Conclusion}

This paper introduced \textbf{EdgeDetect}, a privacy-preserving federated intrusion detection framework designed for resource-constrained 6G-IoT environments. EdgeDetect employs \emph{gradient smartification}, a median-based binarization that compresses local updates to $\{+1,-1\}$, achieving a $32\times$ communication reduction while maintaining convergence. Combined with Paillier homomorphic encryption, the framework ensures that only aggregated updates are revealed to the server, mitigating gradient inversion and honest-but-curious threats. Experiments on CIC-IDS2017 (2.8M flows, 7 attack classes) show that EdgeDetect achieves $98.0\%$ accuracy and $97.9\%$ macro F1, matching centralized performance while reducing per-round communication from $450$~MB to $14$~MB ($96.9\%$ reduction). Ablation analysis confirms that smartification enables efficient compression with negligible utility loss, encryption prevents gradient reconstruction (PSNR $15.1$~dB vs.\ $31.7$~dB undefended), and SMOTE significantly improves minority-class recall. Under $5\%$ poisoning and severe data imbalance, the system maintains $87\%$ accuracy and $0.95$ minority-class F1 ($p<0.001$), demonstrating robustness for real-world deployment. Edge experiments on Raspberry Pi~4 further validate practicality, achieving a $4.2$~MB memory footprint, $0.8$~ms latency, and $12$~mJ per inference with minimal accuracy degradation. Overall, EdgeDetect demonstrates that secure federated IDS can meet the strict privacy, efficiency, and reliability requirements of next-generation 6G-IoT edge networks.

\section*{Acknowledgments}
We thank the Canadian Institute for Cybersecurity for providing the CIC-IDS2017 dataset and the anonymous reviewers for their valuable feedback that improved this work.

\bibliographystyle{IEEEtran}
\bibliography{references}

\appendix
\section{Theoretical Analysis and Gradient Smartification}
\label{appendix:theory}

This appendix provides additional theoretical clarification of the proposed Gradient Smartification mechanism, its convergence properties relative to signSGD, and the adversarial threat model addressed by the combined binarization and encryption framework.

\begin{table}[ht]
\centering
\caption{Hyperparameter configurations (nested 3-fold CV).}
\label{tab:hyperparams}
\setlength{\tabcolsep}{20pt}
\renewcommand{\arraystretch}{1.0}
\small
\begin{tabular}{@{}ll@{}}
\toprule
\textbf{Model} & \textbf{Configuration} \\
\midrule
LogReg & $C\!\in\!\{0.1,100\}$; solver=\{saga,sag\}; $\ell_2$ \\
SVM-RBF & $C\!=\!1.0$; $\gamma\!=\!0.001$ \\
RF & $n\!\in\!\{100,200\}$; depth=20; $m=\lfloor\sqrt{d}\rfloor$ \\
DT & depth$\!\in\!\{6,10,15\}$; min\_split=5 \\
KNN & $k\!\in\!\{3,5,7\}$; wt=\{uniform,dist\} \\
GB & $\nu=0.1$; $n=100$ \\
MLP & [128,64]; drop=0.5; Adam($10^{-3}$) \\
\bottomrule
\end{tabular}
\end{table}

\subsection{Relationship to signSGD}

Classical signSGD updates model parameters using the element-wise sign of stochastic gradients:
\begin{equation}
W^{(r+1)} = W^{(r)} - \eta \cdot \mathrm{sign}(\nabla \mathcal{L}(W^{(r)})).
\end{equation}

In contrast, EdgeDetect applies a median-centered binarization:
\begin{equation}
\Delta_{i,\text{bin}}^{(r)} = \mathrm{sign}\!\left(\Delta_i^{(r)} - \theta\right), 
\quad \theta = \mathrm{median}(|\Delta_i^{(r)}|).
\end{equation}

Unlike signSGD, which thresholds at zero, the proposed formulation suppresses low-magnitude gradient components whose absolute values fall below the median. This reduces stochastic noise and mitigates the influence of small-variance gradient coordinates common in high-dimensional IDS feature spaces.

\subsection{Key Distinction}
Let's $\Delta_i^{(r)} = g + \epsilon$ denote the true gradient $g$ with stochastic noise $\epsilon$. Under zero-threshold binarization, small-noise perturbations may flip signs when $|g_j|$ is small. Median-threshold binarization suppresses coordinates where $|g_j| < \theta$, reducing sign-flip probability and lowering gradient variance. Empirically, this improves convergence stability under heterogeneous client distributions.

\subsection{Convergence Sketch}

Assume: $\mathcal{L}(W)$ is $L$-smooth, Stochastic gradients are unbiased: $\mathbb{E}[\Delta_i^{(r)}] = \nabla \mathcal{L}(W^{(r)})$, Gradient variance is bounded: $\mathbb{E}\|\Delta_i^{(r)} - \nabla \mathcal{L}\|^2 \leq \sigma^2$. Under these conditions, signSGD achieves a convergence rate:
\begin{equation}
\mathbb{E}[\|\nabla \mathcal{L}(W^{(r)})\|^2] = O\!\left(\frac{1}{\sqrt{r}}\right).
\end{equation} 

Since Gradient Smartification preserves dominant gradient directions and discards only low-magnitude coordinates, the update remains directionally aligned with $\nabla \mathcal{L}$ in expectation. Therefore, its convergence behavior asymptotically matches signSGD under bounded noise assumptions. Full formal proof is left for future work; empirical convergence curves in Section~V support this claim.

\subsubsection{Proof Sketch of Theorem 1}

We assume $f$ is $L$-smooth. By standard smoothness inequality,
\[
f(w_{t+1}) 
\le 
f(w_t) 
+ \langle \nabla f(w_t), w_{t+1}-w_t \rangle
+ \frac{L}{2}\|w_{t+1}-w_t\|_2^2.
\]

Substituting the update rule 
$w_{t+1} = w_t - \eta \tilde{g}_t$ gives
\[
f(w_{t+1})
\le
f(w_t)
- \eta \langle \nabla f(w_t), \tilde{g}_t \rangle
+ \frac{L\eta^2}{2}\|\tilde{g}_t\|_2^2.
\]

Taking expectation and applying Proposition 1,
\[
\mathbb{E}\big[\langle \nabla f(w_t), \tilde{g}_t \rangle\big]
\ge
\gamma \|\nabla f(w_t)\|_2^2.
\]

Thus,
\[
\mathbb{E}[f(w_{t+1})]
\le
f(w_t)
- \eta \gamma \|\nabla f(w_t)\|_2^2
+ \frac{L\eta^2}{2} d.
\]

Choosing $\eta = \mathcal{O}(1/\sqrt{T})$ and summing over $T$ iterations yields
\[
\min_{t\le T}
\mathbb{E}\|\nabla f(w_t)\|_2^2
=
\mathcal{O}\!\left(\frac{1}{\gamma \sqrt{T}}\right),
\]
establishing convergence to a stationary point with a degradation factor $1/\gamma$ due to binarization.

\subsection{Bias and Stability of Gradient Smartification}

Median-threshold binarization introduces coordinate-wise bias:
\begin{equation}
\mathbb{E}[\Delta_{\text{bin}}] \neq \nabla \mathcal{L}.
\end{equation}

\subsection{Computational Performance Analysis}
\label{subsec:computational}

Table~\ref{tab:computational} quantifies training time and inference latency across all evaluated models, establishing practical feasibility for real-time intrusion detection deployment.

\begin{table}[ht]
\centering
\caption{Computational Performance: Training Time and Inference Latency}
\label{tab:computational}
\setlength{\tabcolsep}{1pt}
\renewcommand{\arraystretch}{1}
\scriptsize
\begin{tabular}{@{}lcccc@{}}
\toprule
\textbf{Model} & \textbf{Configuration} & \textbf{Train Time (s)} & \textbf{Inference (ms)} & \textbf{Memory (MB)} \\
\midrule
Logistic Reg. & $C=100$, sag & 2.4 & 0.12 & 45 \\
SVM & rbf, $C=1$, $\gamma=0.1$ & 18.7 & 1.45 & 178 \\
Random Forest & $T=15$, $d=8$ & 12.3 & 0.87 & 234 \\
Decision Tree & $d=10$ & 1.1 & 0.08 & 28 \\
KNN & $k=7$, distance-wt & 0.3$^*$ & 3.21 & 412$^\dagger$ \\
\bottomrule
\end{tabular}

\vspace{2mm}
\begin{minipage}{\linewidth}
\footnotesize
\textit{Notes:} Benchmarked on Intel i7-9700K @ 3.6GHz, 32GB RAM, single-threaded execution. Train Time includes hyperparameter search, cross-validation, and final model fitting on the full training set ($n=12{,}000$ for binary; $n=28{,}000$ for multi-class). Inference was measured per sample on the test set. Memory = peak RAM consumption during training. $^*$KNN training is instantaneous (lazy learning) but requires $^\dagger$412 MB to store all training instances for prediction.
\end{minipage}
\end{table}

\subsection{Per-Class Performance Analysis}
\label{subsec:per_class}

Table~\ref{tab:per_class_metrics} reports detailed per-class results for the best Random Forest configuration ($T=15$, depth=8). Errors are asymmetric across attack families, reflecting different separability in the PCA feature space.

\begin{table}[ht]
\centering
\caption{Per-Class Performance Breakdown for Random Forest (Config 2)}
\label{tab:per_class_metrics}
\setlength{\tabcolsep}{2pt}
\renewcommand{\arraystretch}{1}
\footnotesize
\begin{tabular}{@{}lcccccc@{}}
\toprule
\textbf{Class} & \textbf{True Pos.} & \textbf{False Pos.} & \textbf{False Neg.} & \textbf{Precision} & \textbf{Recall} & \textbf{F1} \\
\midrule
BENIGN & 992 & 8 & 15 & 0.992 & 0.985 & \textbf{0.989} \\
DoS & 978 & 12 & 10 & 0.988 & 0.990 & \textbf{0.989} \\
DDoS & 975 & 14 & 11 & 0.986 & 0.989 & \textbf{0.987} \\
Port Scan & 935 & 42 & 23 & 0.957 & 0.976 & \textbf{0.966} \\
Brute Force & 928 & 48 & 24 & 0.951 & 0.975 & \textbf{0.963} \\
Web Attack & 885 & 78 & 37 & 0.919 & 0.960 & \textbf{0.939} \\
Bot & 863 & 94 & 43 & 0.902 & 0.953 & \textbf{0.927} \\
\midrule
\textbf{Macro Avg.} & \textbf{6{,}556} & \textbf{296} & \textbf{163} & \textbf{0.956} & \textbf{0.975} & \textbf{0.966} \\
\bottomrule
\end{tabular}

\vspace{1.5mm}
\begin{minipage}{\linewidth}
\footnotesize
\textit{Notes:} Test set $n=7{,}000$ (1,000 per class). True Pos./False Pos./False Neg.\ are computed one-vs-rest per class. Macro averages weight all classes equally.
\end{minipage}
\end{table}

\subsubsection{Non-IID Data Distribution Analysis}
Table~\ref{tab:noniid_perclass} reports per-class F1 under increasing heterogeneity. Performance degrades smoothly as it $\alpha$ decreases, with minority/overlapping classes most affected.

\begin{table}[ht]
\caption{Per-Class F1-Scores Under Data Heterogeneity}
\label{tab:noniid_perclass}
\centering
\scriptsize
\setlength{\tabcolsep}{6pt}
\begin{tabular}{lcccccc}
\hline
\textbf{Attack Class} & \textbf{IID} & $\alpha$\textbf{=10} & $\alpha$\textbf{=1.0} & $\alpha$\textbf{=0.5} & $\alpha$\textbf{=0.1} & \textbf{Label Skew} \\
\hline
BENIGN & 0.989 & 0.987 & 0.983 & 0.978 & 0.971 & 0.984 \\
DoS & 0.989 & 0.988 & 0.985 & 0.981 & 0.974 & 0.987 \\
DDoS & 0.987 & 0.986 & 0.982 & 0.976 & 0.968 & 0.981 \\
Port Scan & 0.966 & 0.964 & 0.957 & 0.948 & 0.934 & 0.961 \\
Brute Force & 0.963 & 0.961 & 0.952 & 0.941 & 0.923 & 0.956 \\
Web Attack & 0.939 & 0.936 & 0.924 & 0.908 & 0.881 & 0.929 \\
Bot & 0.927 & 0.923 & 0.908 & 0.889 & 0.854 & 0.918 \\
\hline
\textbf{Macro Avg.} & \textbf{0.966} & \textbf{0.964} & \textbf{0.956} & \textbf{0.946} & \textbf{0.929} & \textbf{0.959} \\
\textbf{Accuracy} & \textbf{98.0} & \textbf{97.8} & \textbf{96.8} & \textbf{95.7} & \textbf{94.2} & \textbf{97.1} \\
\hline
\end{tabular}
\begin{flushleft}
\scriptsize
\textit{Notes:} $\alpha$ is the Dirichlet concentration parameter (smaller, $\alpha$ $\Rightarrow$ higher heterogeneity). Label skew assigns each client 2–3 dominant classes with 70\% probability. $K=50$, $C=1.0$, averaged over 5 runs.
\end{flushleft}
\end{table}



\vspace{11pt}


\vspace{11pt}


\vfill

\end{document}